\documentclass[11pt,a4paper]{article}
\usepackage{indentfirst}
\usepackage{color}
\usepackage{amsmath}
\usepackage{amssymb}
\usepackage[utf8]{inputenc}
\usepackage{amsfonts}
\usepackage[header, toc]{appendix}
\numberwithin{equation}{section}

\newcommand{\be}{\begin{equation}}
\newcommand{\ee}{\end{equation}}
\newcommand{\ba}{\begin{eqnarray}}
\newcommand{\ea}{\end{eqnarray}}

\begin{document}



\begin{center}
	
	
	{\Large \bf Four-Four Maxwell Equations}\\	
	{\large R. Doria\footnote{doria@aprendanet.com} and}
	{\large L.S. Mendes \footnote{santiago.petropolis@outlook.com}}\\[0.5cm]
	
	{\large $^1$Aprendanet, Petrópolis, Brazil; Quarks, Petropolis, Brazil}\\
	{\large $^2$Aprendanet, Petrópolis, Brazil; CEFET-RJ, Petropolis, Brazil}
\end{center}

\begin{abstract}

Electromagnetism is the energy originating from an electric charge. Our purpose is to enlarge Maxwell. Include the charge transfer phenomenology. A four bosons electromagnetism is derived. An EM completeness is achieved. The charge's set $\{+,0,-\}$ is  intermediated by a quadruplet $\{A_{\mu}, U_{\mu}, V^{\pm}_{\mu}\}$. They are the usual photon $A_{\mu}$, massive photon $U_{\mu}$ and charged photons $V^{\pm}_{\mu}$.

A new electric charge symmetry arises. It is ruled by an extended abelian symmetry $U_{q}\equiv U(1)\times SO(2)_{global}$.  It generates a Lagrangian and three distinct Noether equations. An EM beyond charge-based emerges. Similarly, as mass is related to Newton laws, electric charge corresponds to the symmetry equation. Its associativity is more than being a coupling constant. It holds a field's flux.

Maxwell is expanded.  Ampère, Faraday, and Gauss laws are improved. Nonlinear granular and collective fields strengths were introduced. Potential fields are explicit. Polarization and magnetization vectors from first principles. New induced Faraday laws. Monopoles depend on fields. Spin-1 and spin-0 sectors separated covariants.

The quadruplet completeness encodes a new EM energy. A journey from Maxwell to photonics is expressed. Seven EM integrated regimes are coursed. Maxwell, systemic, nonlinear, neutral, spintronics, electroweak, photonics. A new relationship for colliding light-light and light-matter appears. Selfinteracting photons are introduced.
\end{abstract}

\section{Introduction}

Electromagnetism is the theory of electric charge, light, and spin. Maxwell electromagnetism is the standard model [1]. However, although all successful, it contains limitations as being linear, subsidiary potential fields, polarization, and magnetization vectors defined by hand, passive light, and heuristic spin. Different subjects such as condensate matter, plasma, and astrophysics are requiring an extension of Maxwell equations. New impacts on light should be considered. Light signals at different frequencies are no longer able to be interpreted by Maxwell linear EM, especially with gamma rays bursts [2] and super-power lasers [3].

The $21^{th}$ century challenge is an EM beyond Faraday-Maxwell [4]. Nonlinearity, strong magnetic fields, and photonics are showing new phenomenologies. A new EM is required. Nevertheless, preserving Maxwell principles as light invariance, electric charge conservation, EM fields in pairs, and EM connected equations. Literature provides 57 models beyond Maxwell were 14 in the Standard Model context, 14 by extensions motivated beyond SM, 19 nonlinear extensions, and 10 due to dimensionality extensions [5]. Most of them are  Euler-Heisenberg and Born-Infeld effective theories [6] and LSV [7] types.

A fundamental EM is expected. A model reinterpreting on an electric charge, light, and spin. Our research considers charge transfer.  There is a microscopic EM that was not registered by Coulomb balance and Maxwell equations to be explored. The elementary particle's electric charge mutation contains an EM to be studied. Analyse the charges set $\{+,0,-\}$ transformations. A physics where charges are created and destroyed, and their energies converted between themselves.

New completeness for EM processes is introduced. Charge transfer extends on EM meaning. A physical context to transmit the electric charge tryad is introduced. The presence of four bosons intermediations is required. However, they do not act isolated, but as an EM set. A four bosons EM to be explored as a whole. A relationship interlaced by electric charge symmetry. 

Thus, a first consideration to look for a fundamental EM comes from electric charge symmetry. An electric charge homotethy is proposed. While homotethy introduces an angle $\theta$ as the mathematical parameter for trigonometry, electric charge provides a gauge parameter $\alpha$ for gauge theory. A performance is constituted by associating four bosons. New EM Lagrangian and Noether identities are expressed. And, corresponding to functions sin$\theta$, cos$\theta$, new EM observables are derived. A meaning of electricity and magnetism beyond Maxwell is discovered.

The second aspect of improving EM is primordial light. The relationship between light and EM is still open. Since Al Haytham light interrogates physics [8]. Maxwell's chief discussion of his electromagnetic theory of light is in his dynamical theory of the electromagnetic field [9]. However, light as an electromagnetic wave contains a contradiction. Concurrently, that light is invariant, it is produced from electric charge oscillations. Arising the discussion: light is a cause or a consequence?

This conundrum is the essence of electromagnetism. However, neither Standard Model nor QED answer that. Following the Big Bang, the fiat lux occurred at $10^{-10} s$, just after the Early Universe second phase transition $SU_{C}(3)\times SU_{L}(2)\times U_{R}(1) \stackrel{SSB}{\rightarrow} SU(3)_{C}(3) \times U_{em}(1)$. For QED, virtual photons are depending on the fine structure constant.  A perspective beyond the electroweak phase transition (EWPT) is necessary [10].

The light metric should work as a framework for searching for such primordial light. It defines space-time symmetries. Historically, light symmetry opened three rooms for physics to be delivered. Maxwell and electric charge [1]; relativity with space-time and matter-energy correlations [11]; Lorentz Group and spin [12]. Nevertheless, the light remained as a consequence. 

Physics contains the mission of discovering a fundamental EM containing the primordial light. Perceive as primordial light properties as invariance, ubiquous, and selfinterating photons. There are various light manifestations beyond Maxwell to be investigated by a Lagrangian. Transcript at tree level a light physics beyond Maxwell. Firstly, develop the $\gamma \gamma$ phenomenology as a revival to Breit-Wheeler scattering [13]. Include others EM manifestations, as magnetogenese [14], light and the Later Universe [15], light at LHC [16], $e^{+}-e^{-}$ in lasers [17], astrophysics [18], $\gamma - Z^{0}$ interaction [19], new light states [20]. 

A fourth light metric consideration will support this investigation. It covers the existence of field families at Lorentz Group. The $\{\frac{1}{2}, \frac{1}{2} \}$ Lorentz Group representation includes the fields set $A_{\mu I} \equiv \{A_{\mu}, U_{\mu}, V_{\mu}^{\pm} \}$ [21]. It supports the four messengers required for conducting the electric charge microscopic flux processes with $\Delta Q = 0, \pm1$. Four photons intermediations will correspond to the most generic microscopic electrodynamics where $A_{\mu}$ means the usual photon, $U_{\mu}$ a massive photon, and $V_{\mu}^{\pm}$ two charged photons. 

A four bosons electromagnetism is generated [22]. It provides a fundamental microscopic electromagnetism. Based on electric charge symmetry, an extended abelian electrodynamics is associated to $U(1)_{q} \equiv U(1)\times SO(2)_{global}$. A Lagrangian with a quadruplet of interconnect fields is obtained and three Noether identities are derived [23]. The third aspect for improving EM is to incorporate spin on fields as $A'_{\mu I} = \left(e^{i\omega_{\alpha \beta}\Sigma^{\alpha \beta}}\right)^{\nu}_{\mu}A_{\nu}$ [24].

A new EM completeness was introduced. It is based on primitive quadruplet potential fields. Literature at most considers the presence of two photons together. It works with the usual photon, massive Proca field, longitudinal, paraphoton, photino [25]. Just the Standard Model propitiates four intermediate bosons together [26]. The four bosons EM includes four photons associated through electric charge symmetry.
 
The objective of this work is to study the vectorial equations corresponding to the four fundamental EM. Starting by writing down the quadri-potentials physics associated by the $U(1)_{q}$ extended abelian symmetry. It gives,

\begin{eqnarray}
	A'_\mu &=& A_\mu + k_1 \partial_\mu \alpha \label{Usual photon}\\ 
	U'_\mu &=& U_\mu + k_2 \partial_\mu \alpha \\
	V_\mu^{+'} &=& e^{iq\alpha} \left( V_\mu^{+} +k_{+} \partial_\mu \alpha\right) \\
	V_\mu^{-'} &=& e^{-iq\alpha} \left( V_\mu^- + k_{-} \partial_\mu \alpha\right). \label{Charge photon}
\end{eqnarray}

Eqs. (1.1-1.4) introduce a potential fields system whose interconnectivity is ruled by the electric charge symmetry. A wholeness is expressed by the parameter $\alpha$. It expresses a new performance beyond Maxwell equations. The so-called four-four Maxwell equations. 

The corresponding antisymmetric vectorial fields strengths are written as

\begin{align}
	F^I_{\mu \nu} &= \partial_{\mu} A^{I}_{\nu}-\partial_{\nu}A^{I}_{\mu};& \vec{E}_{i}^{I}&=F^{I}_{0i}; & \vec{B}_{i}^{I}&=\frac{1}{2}\epsilon_{ijk}F^{I}_{jk} 
\end{align}
where $A^{\mu}_{I} \equiv \left(\phi_{I}, \vec{A}_{I}\right)$ and $I$ is a flavour indice  $I = 1,...4$ corresponding to $A_{\mu}, U_{\mu}, V^{\pm}_{\mu}$

For the granular symmetric fields strengths, one gets
\begin{align}
	S^{I}_{\mu \nu} &=\partial_{\mu}A^{I}_{\nu} + \partial_{\nu}A^{I}_{\mu};& S^{\alpha I}_{\alpha}& =2\partial_{\alpha}A^{\alpha I}
\end{align}
\begin{align}
	S^{I}_{0i} &= \partial_{0}A^{I}_{i} + \partial_{i}A^{I}_0, & S^{I}_{ij} &= \partial_{i}A^{I}_{j} + \partial_{j}A^{I}_{i}
\end{align}

The antisymmetric collective fields strengths are written as
\begin{align}
	e_{[\mu \nu]} &= \mathbf{e}_{[IJ]}A^{I}_{ \mu}A^{J}_\nu,& \vec{\mathbf{e}}_{i}&=\mathbf{e}_{[0i]},& \vec{\mathbf{b}}_{i} &= \frac{1}{2}\epsilon_{ijk}\mathbf{e}_{[jk]}
\end{align}


It yields the following antisymmetric group of collective vectors fields:
\begin{align}
	\vec{\mathbf{e}}_{AU} &= \mathbf{e}^{[0i]}_{AU} = \mathbf{e}_{[12]}\left(\phi_{A}\vec{U}-\phi_{U}\vec{A}\right) & \vec{b}_{AU} &= \mathbf{e}^{[ij]}_{AU} = \mathbf{e}_{[12]}\left(\vec{A}\times \vec{U}\right)
	\\
	\vec{\mathbf{e}}_{+-} &= \mathbf{e}^{[0i]}_{+-} = \mathbf{e}_{[34]}\left(\phi_{-}\vec{V}^+-\phi_{+}\vec{V}^{-}\right)& \vec{b}_{+-} &= \mathbf{e}^{[ij]}_{+-}= \mathbf{e}_{[34]}\left(\vec{V}^+ \times \vec{V}^{-}\right)
	\\
	\vec{\mathbf{e}}_{+A} &= \mathbf{e}^{[0i]}_{+A}\left(\mathbf{e}_{[13]}+i\mathbf{e}_{[14]}\right)\left(\phi_{A}\vec{V}^{+}-\phi_{+}\vec{A}\right)&   \vec{b}_{+A}&= \mathbf{e}^{[ij]}_{+A}=\left(\mathbf{e}_{[13]}+i\mathbf{e}_{[14]}\right)\left(\vec{A}\times\vec{V}^+\right)
	\\
	\vec{\mathbf{e}}_{+A} &= \mathbf{e}^{[0i]}_{-A}\left(\mathbf{e}_{[13]}-i\mathbf{e}_{[14]}\right)\left(\phi_{A}\vec{V}^{-}-\phi_{-}\vec{A}\right)&   \vec{b}_{-A}&= \mathbf{e}^{[ij]}_{-A}=\left(\mathbf{e}_{[13]}-i\mathbf{e}_{[14]}\right)\left(\vec{A}\times\vec{V}^-\right)
	\\
	\vec{\mathbf{e}}_{+U} &= \mathbf{e}^{[0i]}_{+U}\left(\mathbf{e}_{[23]}+i\mathbf{e}_{[24]}\right)\left(\phi_{U}\vec{V}^{+}-\phi_{+}\vec{U}\right)&   \vec{b}_{+U}&= \mathbf{e}^{[ij]}_{+U}=\left(\mathbf{e}_{[23]}+i\mathbf{e}_{[24]}\right)\left(\vec{U}\times\vec{V}^+\right)
	\\
	\vec{\mathbf{e}}_{-U} &= \mathbf{e}^{[0i]}_{-U}\left(\mathbf{e}_{[23]}-i\mathbf{e}_{[24]}\right)\left(\phi_{U}\vec{V}^{-}-\phi_{-}\vec{U}\right)&   \vec{b}_{-U}&= \mathbf{e}^{[ij]}_{+U}=\left(\mathbf{e}_{[23]}+i\mathbf{e}_{[24]}\right)\left(\vec{U}\times\vec{V}^+\right)
\end{align}

For symmetric collective vector fields:
\begin{align}
	s^{\alpha}_{\alpha AA} &= \mathbf{e}_{(11)}A^{\alpha}A_{\alpha}, & s^{\alpha}_{\alpha UU} &= \mathbf{e}_{(22)}U^{\alpha}U_{\alpha} & s^{\alpha}_{\alpha AU}&=\mathbf{e}_{(12)}A^{\alpha}U_{\alpha} \\
	\vec{s}_{AA}& = \mathbf{e}^{(0i)}_{(AA)} = \mathbf{e}_{(11)}\phi_{A}\vec{A} & \vec{s}& = \mathbf{e}^{(0i)}_{UU} = \mathbf{e}_{(22)}\phi_{U}\vec{U} & \vec{s}_{AU} = \mathbf{e}^{(0i)}_{AU} &= \mathbf{e}_{(12)}\left(\phi_{U}\vec{A} + \phi_{A}\vec{U}\right)
\end{align}
and
\begin{align}
	s^{\alpha}_{\alpha +-}& =\left(\mathbf{e}_{(33)}+\mathbf{e}_{(44)}\right)V^{ + \alpha }V^{-}_{\alpha} & \vec{\mathbf{s}}_{+-}&= \left(\mathbf{e}_{(33)}+\mathbf{e}_{(44)}\right)\left(\phi_{-}\vec{V}^{+}+\phi_{+}\vec{V}^{-}\right)\\
	s^{\alpha}_{\alpha ++}& = \left(\mathbf{e}_{(33)}-\mathbf{e}_{(44)}\right)V^{+\alpha}V_{+\alpha}& \vec{s}_{++}&=\mathbf{e}^{(0i)}_{++} = \left(\mathbf{e}_{(33)}-\mathbf{e}_{(44)}\right)\phi_{+}\vec{V}^{+}\\
	s^{\alpha}_{\alpha --}& = \left(\mathbf{e}_{(33)}-\mathbf{e}_{(44)}\right)V^{-\alpha}V_{-\alpha}& \vec{s}_{--}&=\mathbf{e}^{(0i)}_{--} = \left(\mathbf{e}_{(33)}-\mathbf{e}_{(44)}\right)\phi_{+}\vec{V}^{-}\\
	s^{\alpha}_{\alpha A+}&= \left(\mathbf{e}_{(13)} + i\mathbf{e}_{(14)}\right)A_{\alpha}V^{+ \alpha}& \vec{s}_{+A}&=\mathbf{e}^{0i}_{A+} =\left(\mathbf{e}_{(13)} + i\mathbf{e}_{(14)}\right)\left(\phi_{+}\vec{A} + \phi_{A}\vec{V}^{+}\right) \\
	s^{\alpha}_{\alpha A-}&= \left(\mathbf{e}_{(13)} - i\mathbf{e}_{(14)}\right)A_{\alpha}V^{- \alpha}& \vec{s}_{-A}&=\mathbf{e}^{0i}_{A-} =\left(\mathbf{e}_{(13)} - i\mathbf{e}_{(14)}\right)\left(\phi_{-}\vec{A} + \phi_{A}\vec{V}^{-}\right) \\
	s^{\alpha}_{\alpha U+}&= \left(\mathbf{e}_{(23)} + i\mathbf{e}_{(24)}\right)U_{\alpha}V^{+ \alpha}& \vec{s}_{+U}&=\mathbf{e}^{0i}_{U+} =\left(\mathbf{e}_{(23)} + i\mathbf{e}_{(24)}\right)\left(\phi_{+}\vec{U} + \phi_{U}\vec{V}^{+}\right) \\
	s^{\alpha}_{\alpha U-}&= \left(\mathbf{e}_{(23)} - i\mathbf{e}_{(24)}\right)U_{\alpha}V^{- \alpha}& \vec{s}_{U-}&=\mathbf{e}^{0i}_{U-} =\left(\mathbf{e}_{(23)} + i\mathbf{e}_{(24)}\right)\left(\phi_{-}\vec{U} + \phi_{U}\vec{V}^{-}\right)
\end{align}
and
\begin{eqnarray}
	\stackrel{\leftrightarrow}{s}_{+-} & =&s^{ij}_{+-} = \left(\mathbf{e}_{(33)}+ \mathbf{e}_{(44)}\right)\left(V^{+i}V^{-j}+V^{-i}V^{+j}\right) 
	\\
	\stackrel{\leftrightarrow}{s}_{AU} & =&s^{ij}_{AU} = \mathbf{e}_{(12)}\left(A^{i}U^{j}+U^{i}A^{j}\right)
	\\
	\stackrel{\leftrightarrow}{s}_{++} & =&s^{ij}_{++} = \left(\mathbf{e}_{(33)}- \mathbf{e}_{(44)}\right)V^{+i}V^{+j} 
	\\
	\stackrel{\leftrightarrow}{s}_{--} & =s&^{ij}_{--} = \left(\mathbf{e}_{(33)}- \mathbf{e}_{(44)}\right)\left(V^{-i}V^{-j}+V^{-i}V^{-j}\right)
	\\
	\stackrel{\leftrightarrow}{s}_{A+} & =&s^{ij}_{A+} = \left(\mathbf{e}_{(13)}+ \mathbf{e}_{(14)}\right)\left(A^{i}V^{+j}+V^{+i}A^{j}\right) 
	\\
	 \stackrel{\leftrightarrow}{s}_{-A} & =&s^{ij}_{--} = \left(\mathbf{e}_{(13)}- \mathbf{e}_{(14)}\right)\left(A^{i}V^{-j}+V^{-i}A^{j}\right)
	\\
	\stackrel{\leftrightarrow}{s}_{U+} & =&s^{ij}_{U+} = \left(\mathbf{e}_{(23)}- \mathbf{e}_{(24)}\right)\left(U^{i}V^{+j}+V^{+i}A^{j}\right) 
	\\
	\stackrel{\leftrightarrow}{s}_{U-} & =&s^{ij}_{U-} = \left(\mathbf{e}_{(23)}- \mathbf{e}_{(24)}\right)\left(U^{i}V^{-j}+V^{-i}U^{j}\right)
\end{eqnarray}

Eqs (1.5)-(1.27) are expressing a new perspective on EM energy. New electromagnetic observables are considered. Electricity and magnetism become a more complete set when granular and collective fields are introduced. Notice that $\mathbf{e}_{[12]}, ..., \mathbf{e}_{(24)}$ are parameters expressed in terms of theory-free coefficients. They can take any value without violating the gauge symmetry. Depending on their relationship, gauge-invariant properties are derived. Gauge invariance is proved either through the Lagrangian [27] or by the individual field's strengths [23]. And so, given these physical entities, we should study their corresponding dynamical theory.

\section{Euler-Lagrange and Bianchi Equations}

For the last 150 years, Maxwell's equations have been the dominant electromagnetic theory. QED supplemented with $1\%$ of deviations due to quantum fluctuations. However, they were not enough to produce features such as nonlinear EM, physical potential fields, origin for polarization and magnetization, light carrying its own EM fields, and spin incorporated in the fields. An enlargement to EM becomes necessary.

A four bosons electromagnetic Lagrangian is constituted through charge transfer [22-24]. The associated electric charge tryad  $\{+,0,-\}$ exchanges introduces an EM under a fields set $\{A_{\mu}, U_{\mu}, V^{\pm}_{\mu}\}$. The corresponding Gauss, Amp$\grave{e}$re, Faraday laws are extended. It yields,
\\
\\

For $A^{T}_{\mu}$ (spin-1):
\begin{equation}\label{Gauss equations for A}
	\vec{\nabla} \cdot \left[4a_1 \vec{E_A} + 2 b_1\left(\vec{\mathbf{e}}_{AU}+ \vec{\mathbf{e}}_{+-} \right)\right] = \rho^T_A
\end{equation}
with
\begin{multline}\label{Densitity Current A}
	\rho^T_A \equiv -2\mathbf{e}_{[12]}\left(\vec{E}_A + \vec{E}_U\right)\cdot\vec{U} - \sqrt{2}\left(\mathbf{e}_{[13]}-i\mathbf{e}_{[14]}\right)\left(\vec{E}_+ + \vec{\mathbf{e}}_{+A} + \vec{\mathbf{e}}_{+U} \right)\cdot \vec{V}^- +\\
	-\sqrt{2}\left(\mathbf{e}_{[13]}+i\mathbf{e}_{[14]}\right)\left(\vec{E}_- + \vec{\mathbf{e}}_{-A} + \vec{\mathbf{e}}_{-U}\right) \cdot \vec{V}^+ 
	- 4\left(\mathbf{e}_{[12]}\vec{\mathbf{e}}_{AU}  + \mathbf{e}_{[34]}\vec{\mathbf{e}}_{+-}\right)\cdot \vec{U}\\
\end{multline}
and
\begin{equation}\label{Ampere-maxwell for A}
	\vec{\nabla} \times \left[\vec{B}_A + 2b_1\left(\vec{b}_{AU} + \vec{b}_{+-}\right)\right] - \frac{\partial}{\partial t}\left[\vec{E}_A + 2 b_1\left(\vec{\mathbf{e}}_{AU}+ \vec{\mathbf{e}}_{+-} \right) \right] = \vec{J}^{T}_A
\end{equation}
with
\begin{eqnarray}
	&&\vec{J}^T_A \equiv -2 \mathbf{e}_{[12]}\left[\left(\vec{E}_A + \vec{E}_U + 2\vec{\mathbf{e}}_{AU}\right)\phi_U + \left(\vec{B}_A + \vec{B}_U + \vec{b}_{AU}\right) \times \vec{U} \right] +\nonumber
	\\
	&&- \sqrt{2}\left(\mathbf{e}_{[13]}-i\mathbf{e}_{[14]}\right)\big[\left(\vec{E}_+ + \vec{ \mathbf{e}}_{+A} +\vec{\mathbf{e}}_{+U}\right) \phi^- + \big(\vec{B}_+ + \vec{b}_{+A} + 
	\\
	&&+\vec{b}_{+U}\big) \times \vec{V}^-\big] - \sqrt{2}\left(\mathbf{e}_{[13]}+i\mathbf{e}_{[14]}\right)\big[\left(\vec{E}_- + \vec{\mathbf{e}}_{-A} +\vec{\mathbf{e}}_{-U} \right) \phi^+ 
	\\
	&&+ \left(\vec{B}_- + \vec{b}_{-A} + \vec{b}_{-U} \right) \times \vec{V}^+\big]\nonumber
\end{eqnarray}
with the following conservation law. 

\begin{equation}
	\frac{\partial \rho^{T}_{A}}{\partial t} + \vec{\nabla}\cdot\vec{j}^{T}_{A} = 0
\end{equation}

The corresponding granular Bianchi identity is 
\begin{equation}
	\vec{\nabla} \cdot \vec{B}_A = 0, \qquad \vec{\nabla} \times \vec{E}_A = -\frac{\partial\vec{B}_A}{\partial t}
\end{equation}

Eqs. (2.1-2.6) are introducing a new photon physics. Different from Maxwell, it is no more an electric charge consequence. It produces its own EM field. It yields dynamics with photon fields producing granular and collective EM fields and interconnected to other three intermediate bosons. As predicted by Sch winger critical fields, $E_{c} = \frac{m_{e}^2c^3}{eh}$, $B_{c} = \frac{m_{e}^2c^2}{eh}$, nonlinearity is incorporated. A photon coupling without requiring the presence of an electric charge appears.

Other associated equations are: 
\\
\\

For $U^{T}_{\mu}$: the associated equations are
\begin{equation}\label{Gauss equation for U}
	\vec{\nabla}\cdot \left[4a_2 \vec{E}_U + 2b_2\left(\vec{\mathbf{e}}_{AU}+ \vec{\mathbf{e}}_{+-} \right) \right] = \rho^T_{U}
\end{equation}
and
\begin{equation}\label{Ampere-Maxwell For U}
	\vec{\nabla} \times \left[\vec{B}_U + 2b_2\left(\vec{b}_{AU} + \vec{b}_{+-}\right)\right] - \frac{\partial}{\partial t}\left[\vec{E}_U + 2 b_2\left(\vec{\mathbf{e}}_{AU}+ \vec{\mathbf{e}}_{+-} \right) \right] = \vec{J}^{T}_U
\end{equation}
Where the charges and current densities are expressed in Appendix A.

The corresponding granular Bianchi identity is 
\begin{equation}
	\vec{\nabla} \cdot \vec{B}_U = 0, \qquad \vec{\nabla} \times \vec{E}_U = -\frac{\partial\vec{B}_U}{\partial t}
\end{equation}

For $V^{T+}_{\mu}$: it yields,
\begin{equation}\label{Gauss equation V+}
	\vec{\nabla} \cdot \left[2a_3 \vec{E_+} + 2 b_3\left(\vec{\mathbf{e}}_{A+}+ \vec{\mathbf{e}}_{U+} \right)\right] = \rho^T_+
\end{equation}
and
\begin{equation}\label{Ampere-Maxwell for V+}
	\vec{\nabla} \times \left[\vec{B}_+ + 2b_3\left(\vec{b}_{+A} + \vec{b}_{+U}\right)\right] - \frac{\partial}{\partial t}\left[\vec{E}_+ + 2 b_3\left(\vec{\mathbf{e}}_{+A}+ \vec{\mathbf{e}}_{+U} \right) \right] = \vec{J}^{T}_+	
\end{equation}


The associated granular Bianchi identity is 
\begin{equation}
	\vec{\nabla} \cdot \vec{B}_+ = 0, \qquad \vec{\nabla} \times \vec{E}_+ = -\frac{\partial\vec{B}_+}{\partial t}
\end{equation}

For $V^{T-}_{\mu}$: one gets,
\begin{equation}\label{Gauss equation for V-}
	\vec{\nabla} \cdot \left[2a_3 \vec{E}_- + 2 b_3\left(\vec{\mathbf{e}}_{A-}+ \vec{\mathbf{e}}_{U-} \right)\right] = \rho^T_-
\end{equation}

and
\begin{equation}\label{Ampere-Maxwell for V-}
	\vec{\nabla} \times \left[\vec{B}_- + 2b_3\left(\vec{b}_{-A} + \vec{b}_{-U}\right)\right] - \frac{\partial}{\partial t}\left[\vec{E}_- + 2 b_3\left(\vec{\mathbf{e}}_{-A}+ \vec{\mathbf{e}}_{-U} \right) \right] = \vec{J}^{T}_-	
\end{equation}


The corresponding Bianchi identity is
\begin{equation}
	\vec{\nabla} \cdot \vec{B}_- = 0, \qquad \vec{\nabla} \times \vec{E}_- = -\frac{\partial\vec{B}_-}{\partial t}
\end{equation}

Eqs. (2.7-2.20) will follow the same conservation law as eq. (2.5). The correspondent charges and currents expressions are related at Appendix A.

Thus, eqs. (2.1)-(2.20) are expressing new EM vectorial equations. Four interconnected photons work as their own sources. Proposing granular and collective EM fields (collective fields identified with polarization and magnetization vectors), nonlinearity, and potential fields interacting with EM fields. Equations preserving rotational and translation symmetries [28] and covariant under $A'_{\mu I} = \Lambda_{\mu}^{\nu}A_{\nu I}$ where $\Lambda^{\nu}_{\mu}$ is the Lorentz transformation matrix [29]. 
\\
\\

Considering the spin-0 sector, one gets the corresponding scalar Gauss and Amp$\grave{e}$re laws. It gives, 
\\
\\

For $A^{L}_{\mu}$ (spin-0): 
\begin{eqnarray}\label{spin-0 spatial equation A}
	&&\frac{\partial}{\partial t}[4\left( \beta_1 + \beta_1 \rho_1 + 11\rho_1 \right)S^{\alpha}_{\alpha A} + 2\left(\rho_1\beta_2 + \rho_2 \beta_1 + 22 \rho_1 \rho_2\right) S^{\alpha}_{\alpha U} +  s^{\alpha}_{\alpha AA} +\nonumber
	\\
	&&+2\left(17 \rho_1 - \beta_1 \right)(s^{\alpha}_{\alpha UU} + s^{\alpha}_{\alpha AU} + s^{\alpha}_{\alpha +-})] = \rho^{s}_A
\end{eqnarray}
with
\begin{eqnarray}
	&&\rho^{s}_{A} \equiv 2 \left(\mathbf{e}_{(11)} + \mathbf{e}_{(12)}\right) \big\{ \big(\beta_1\vec{S}_A + \beta_2 \vec{S}_U \big) \cdot \left(\vec{A} + \vec{U}\right) \nonumber 
	\\
	 &&+ \big[\left( \beta_1 + 17\rho_1 \right)S^{\alpha}_{\alpha A} + \left(\beta_2 +17\rho_2\right) S^{\alpha}_{\alpha U} \big] \left(\phi_A + \phi_U\right) \big\} +\nonumber
	\\
	&&+\sqrt{2}\left(\mathbf{e}_{(13)} - i\mathbf{e}_{(14)}\right)\big[\beta_+ \vec{S}^+\vec{V}^- +\left(\beta_+ + 17\rho_+\right)S^{\alpha}_{\alpha +}\phi^-\big] \nonumber
	\\
	&&- \beta_1\mathbf{e}_{(11)}\left(S^{\alpha}_{\alpha A} \phi_A + 2 \vec{S}_A \cdot \vec{A}\right)+\beta_1\mathbf{e}_{(12)}\big(S^{\alpha}_{\alpha A}\phi_U + S^{\alpha}_{\alpha U}\phi_A + \vec{S}_A \cdot \vec{U}  \nonumber
	\\
	&&+ \vec{S}_U \cdot \vec{A}\big) - \beta_1\left(\mathbf{e}_{(33)} + \mathbf{e}_{(44)}\right) \left(S^{\alpha}_{\alpha +}\phi^- + S^{\alpha}_{\alpha -}\phi^+ + \vec{S}_+\vec{V}^- + \vec{S}_-\vec{V}^+\right)\nonumber
	\\
	&&+\sqrt{2}\left(\mathbf{e}_{(13)} + i\mathbf{e}_{(14)}\right)\left[\beta_- \vec{S}^-\vec{V}^+ + \left(\beta_- + 17\rho_-\right)S^{\alpha}_{\alpha -}\phi^+\right] \nonumber
	\\
	&&+ 4\mathbf{e}_{(11)}\left(\vec{s}_{AA} + \vec{s}_{UU} + \vec{s}_{AU} +\vec{s}_{+-} \right) \cdot \vec{A} +72\mathbf{e}_{(11)}\big(s^{\alpha}_{\alpha UU}\nonumber
	\\
	&& + s^{\alpha}_{\alpha AU} + s^{\alpha}_{\alpha+-}\big)\phi_A + 4 \mathbf{e}_{(12)} \left( \vec{s}_{AA} + \vec{s}_{UU} \right) \cdot \vec{U} 
	\\
	&&+ 72 \mathbf{e}_{(12)}\left(s^{\alpha}_{\alpha AA} + \mathbf{e}^{\alpha}_{\alpha UU} + s^{\alpha}_{+-}\right)\phi_U-\beta_1\mathbf{e}_{(22)}\left(\vec{S}_U\cdot \vec{U} + S^{\alpha}_{\alpha U} \right) \nonumber
	\\
	&&+\sqrt{2}\left(\mathbf{e}_{(13)} + i\mathbf{e}_{(14)}\right) \left[\vec{s}_{-A}\vec{V}^+ + 18 s^{\alpha}_{\alpha A-} \phi^+\right] + \nonumber
	\\
	&&\sqrt{2}\left(\mathbf{e}_{(13)} - i\mathbf{e}_{(14)}\right) \left[\vec{s}_{+A}\vec{V}^- + 18 s^{\alpha}_{\alpha A+} \phi^-\right]\nonumber
	\\
	&&+\sqrt{2}\left(\mathbf{e}_{(23)} + i\mathbf{e}_{(24)}\right) \left[\vec{s}_{-U}\vec{V}^+ + 18 s^{\alpha}_{\alpha U-} \phi^+\right] +\nonumber
	\\
	&&+ \sqrt{2}\left(\mathbf{e}_{(23)} - i\mathbf{e}_{(24)}\right) \left[\vec{s}_{+U}\vec{V}^- + 18 s^{\alpha}_{\alpha U+} \phi^-\right]\nonumber
\end{eqnarray}
and the corresponding scalar-Amp$\grave{e}$re law
\begin{eqnarray}
	&&\vec{\nabla}[4\left( \beta_1 + \beta_1 \rho_1 + 11\rho_1 \right)S^{\alpha}_{\alpha A} + 2\left(\rho_1\beta_2 + \rho_2 \beta_1 + 22 \rho_1 \rho_2\right) S^{\alpha}_{\alpha U} + \nonumber
	\\
	&&+2\left(17 \rho_1 - \beta_1 \right) ( s^{\alpha}_{\alpha AA} + s^{\alpha}_{\alpha UU} + s^{\alpha}_{\alpha AU} + s^{\alpha}_{\alpha +-})] =- \vec{j}^{s}_A
\end{eqnarray}
with
\begin{eqnarray}
	&&\vec{j}^{s}_A \equiv 2\beta_1\big\{S^{i0}_A\left(\mathbf{e}_{(11)}A_0 + \mathbf{e}_{(12)}U_0\right) + S^{ij}_A\left( \mathbf{e}_{(11)}A_j + \mathbf{e}_{(12)}U_j \right) \nonumber
	\\
	&&+ S^{\alpha}_{\alpha A}\left( \mathbf{e}_{(11)}A^i + \mathbf{e}_{(12)}U^i \right)\big\} +2\beta_2\big\{S^{i0}_U\left(\mathbf{e}_{(11)}A_0 + \mathbf{e}_{(12)}U_0\right) \nonumber
	\\
	&& + S^{ij}_U\left( \mathbf{e}_{(11)}A_j + \mathbf{e}_{(12)}U_j \right) + S^{\alpha}_{\alpha U}\left( \mathbf{e}_{(11)}A^i + \mathbf{e}_{(12)}U^i \right)\big\}\nonumber
	\\
	&&34\rho_1S^{\alpha}_{\alpha A}\left(\mathbf{e}_{(11)}A^i + \mathbf{e}_{(12)}U^i\right) + 34\rho_2S^{\alpha}_{\alpha U}\left(\mathbf{e}_{(11)}A^i + \mathbf{e}_{(12)}U^i\right) \nonumber
	\\ 
	&&+ \sqrt{2}\left(\mathbf{e}_{(13)} - i\mathbf{e}_{(14)}\right)[\beta_+S^{i0}_+ V_0^- +\beta_+S^{ij}_+V_j^- \left(\beta_+ + 17\rho_+\right)S^{\alpha}_{\alpha +}V^{i-}] + \nonumber
	\\
	&&+\sqrt{2}\left(\mathbf{e}_{(13)} + i\mathbf{e}_{(14)}\right)[\beta_-S^{i0}_- V_0^+ +\beta_-S^{ij}_-V_j^+ \left(\beta_- + 17\rho_-\right)S^{\alpha}_{\alpha -}V^{i+}]+\nonumber
	\\
	&&+4\mathbf{e}_{(11)}[\left(s^{i0}_{AA} + s^{i0}_{UU} + s^{i0}_{AU} + s^{i0}_{+-} \right)A_0 + \left(s^{ij}_{AA} + s^{ij}_{UU} + s^{ij}_{AU} + s^{ij}_{+-} \right)A_j \nonumber
	\\
	&&+ (2s^{\alpha}_{\alpha AA} + 17s^{\alpha}_{\alpha UU} + 17s^{\alpha}_{\alpha UU}+ s^{\alpha}_{\alpha AU} + s^{\alpha}_{\alpha +-})A^i] + 4\mathbf{e}_{(12)}[\big(s^{i0}_{AA} + s^{i0}_{UU} \nonumber
	\\
	&& + s^{i0}_{AU} \big)U_0 + \big(s^{ij}_{AA} + s^{ij}_{UU} + s^{ij}_{AU} \big)U_j + 18(s^{\alpha}_{\alpha AA} + s^{\alpha}_{\alpha UU} + s^{\alpha}_{\alpha AU} + s^{\alpha}_{\alpha +-})U^i] \nonumber
	\\
	&&+ \sqrt{2}\left(\mathbf{e}_{(13)} + i\mathbf{e}_{(14)}\right)\left(s^{i0}_{-A}V^+_0 + s^{ij}_{-A}V^+_j + 18s^{\alpha}_{\alpha -A}V^{i+}\right) + 
	\\
	&&+\sqrt{2}\left(\mathbf{e}_{(13)} - i\mathbf{e}_{(14)}\right)(s^{i0}_{+A}V^-_0 + s^{ij}_{+A}V^-_j 18s^{\alpha}_{\alpha +A}V^{i-}) 	\nonumber
	\\
	&& + \sqrt{2}\left(\mathbf{e}_{(23)} + i\mathbf{e}_{(24)}\right)\left(s^{i0}_{-U}V^+_0 + s^{ij}_{-U}V^+_j + 18s^{\alpha}_{\alpha -U}V^{i+}\right) + \nonumber
	\\
	&&\sqrt{2}\left(\mathbf{e}_{(23)} -i\mathbf{e}_{(24)}\right)(s^{i0}_{+U}V^-_0++s^{ij}_{+U}V^-_j + 18s^{\alpha}_{\alpha +U}V^{i-}) \nonumber
	\\
	&&-\beta_1\left[\mathbf{e}_{(11)}\left(S^{\alpha}_{\alpha A}A^i + 2S^{i0}_A A_0 + S^{ij}_A A_j\right) + \mathbf{e}_{(22)}\left(S^{\alpha}_{\alpha U}U^i + 2S^{i0}_U U_0 +  S^{ij}_U U_j\right)\right]\nonumber
	\\
	&&-\beta_1\mathbf{e}_{(12)}\left[S^{\alpha}_{\alpha U}A^i +S^{\alpha}_{\alpha A}U^i + 2S^{i0}_U A_0 + 2S^{i0}_A U_0 + 2S^{ij}_U A_j + 2 S^{ij}_U A_j \right]\nonumber
	\\
	&&\beta_1\left(\mathbf{e}_{(33)} + \mathbf{e}_{(44)}\right)\big(S^{\alpha}_{\alpha + }V^{i -} + S^{\alpha}_{\alpha -}V^{i+} + 2S^{i0}_+ V^-_0 \nonumber
	\\
	 &&+ 2S^{i0}_- V^+_0 + 2S^{ij}_+ V^-_j + 2S^{ij}_- V^+_j\big)\nonumber
\end{eqnarray}

Notice that due to operating on scalar field strengths, the above longitudinal equations are not related to divergence and rotational. Their field variations are depending just on time and space. Eq. (2.21) is expressing a kind of scalar fields strengths velocity and eq. (2.23) a spatial evolution. The corresponding conservation law is

\begin{equation}
	\nabla \rho^{s}_{A} + \frac{\partial j^{s}_{A}}{\partial t} = 0
\end{equation}
\\
\\

For $U^{L}_{\mu}$:
\begin{eqnarray}
	&&\frac{\partial}{\partial t}[4\left( \beta_2 + \beta_2 \rho_2 + 11\rho_1 \right)S^{\alpha}_{\alpha U} + 2\left(\rho_1\beta_2 + \rho_2 \beta_1 + 22 \rho_1 \rho_2\right) S^{\alpha}_{\alpha A} \nonumber
	\\
	&& 2\left(17 \rho_2 - \beta_2 \right) ( s^{\alpha}_{\alpha AA} +s^{\alpha}_{\alpha UU} + s^{\alpha}_{\alpha AU} + s^{\alpha}_{\alpha +-})] - 2\mathbf{m}^2_U\phi_U = \rho^{s}_U
\end{eqnarray}
and
\begin{eqnarray}
	&&\vec{\nabla}[4\left( \beta_2 + \beta_2 \rho_2 + 11\rho_2 \right)S^{\alpha}_{\alpha U} + 2\left(\rho_1\beta_2 + \rho_2 \beta_1 + 22 \rho_1 \rho_2\right) S^{\alpha}_{\alpha A} \nonumber
	\\
	&&+ 2\left(17 \rho_2 - \beta_2 \right) ( s^{\alpha}_{\alpha AA} +s^{\alpha}_{\alpha UU} + s^{\alpha}_{\alpha AU} + s^{\alpha}_{\alpha +-})] 2\mathbf{m}^2_U\vec{U} = -\vec{j}^{s}_U
\end{eqnarray}

For $V^{L-}_{\mu}$:
\begin{eqnarray}
	&&\frac{\partial}{\partial t}\left\{2\left(\beta_+\beta_- + 16\rho_+\rho_- + \rho_+\beta_- + \rho_+ \beta_+\right)S^{\alpha}_{\alpha -} + \left(34\rho_+ - 2\beta_-\right)\left(s^{\alpha}_{\alpha -A} + s^{\alpha}_{\alpha -U}\right)\right\}\nonumber
	\\
	&&- \mathbf{m}^2_{V}\phi^- = \rho^{-s}_{V}
\end{eqnarray}
and
\begin{eqnarray}
	&&\vec{\nabla}\left\{2\left(\beta_+\beta_- + 16\rho_+\rho_- + \rho_+\beta_- + \rho_+ \beta_+\right)S^{\alpha}_{\alpha -} + \left(34\rho_+ - 2\beta_-\right)\left(s^{\alpha}_{\alpha -A} + s^{\alpha}_{\alpha -U}\right)\right\}\nonumber 
	\\
	&&- \mathbf{m}^2_{V}\vec{V}^- = -\vec{j}^{-s}_{V}
\end{eqnarray}

For $V_{\mu}^{L+}$:
\begin{eqnarray}
	&&\frac{\partial}{\partial t}\left\{2\left(\beta_+\beta_- + 16\rho_+\rho_- + \rho_+\beta_- + \rho_+ \beta_+\right)S^{\alpha}_{\alpha +} + \left(34\rho_+ - 2\beta_+\right)\left(s^{\alpha}_{\alpha +A} + s^{\alpha}_{\alpha +U}\right)\right\}\nonumber
	\\
	 &&- \mathbf{m}^2_{V}\phi^+ = \rho^{+s}_{V}
\end{eqnarray}
and
\begin{eqnarray}
	&&\vec{\nabla}\left\{2\left(\beta_+\beta_- + 16\rho_+\rho_- + \rho_+\beta_- + \rho_+ \beta_+\right)S^{\alpha}_{\alpha +} + \left(34\rho_+ - 2\beta_+\right)\left(s^{\alpha}_{\alpha +A} + s^{\alpha}_{\alpha +U}\right)\right\}\nonumber 
	\\
	&&- \mathbf{m}^2_{V}\vec{V}^+ = -\vec{j}^{+s}_{V}
\end{eqnarray}
where the above charges and current densities are written at Apendice A.

Concluding, these Euler-Lagrange equations, as Maxwell equations, make an interconnected system. Enlarging the EM properties with new electric and magnetic field relationships. Lorentz's symmetry also relates these spatial and temporal equations according to covariance.

\section{Electric Charge Homothety}

Charge transfer $\{+,0,-\}$ introduces a generic electric charge conservation. While Maxwell constructed charge conservation by introducing the displacement current, and as consequence, emerges the connections of the electric and magnetic field, here, a new physicality appears. Maxwell abelian gauge symmetry is extended for $U_{q}(1) \equiv U(1)\times SO(2)_{global}$. It yields the four bosons EM Homothety introduces an electric charge symmetry associated with a kind of homothety where the correspondent angle is given by the gauge parameter $\alpha$. Its symmetry acts at eqs (1.1-1.4) gauge transformations and Noether theorem [30]. 

Noether theorem produces three independent equations. 

\begin{equation}
	\alpha\partial_{\mu}J^{\mu} + \partial_{\nu}\alpha\left\{\partial_{\mu}K^{\mu \nu} + J^{\nu}\right\} + \partial_{\mu}\partial_{\nu}\alpha K^{\mu \nu}=0
\end{equation}
Eqs. (3.1) are called electric charge conservation, symmetry equation, and constraint. While at QED the first two are equal, a new aspect appears through the EM quadruplet. The electric charge equation is identified as the symmetry equation. And so, as inertial mass is characterized by second Newton's law, electric charge defines its own equation in terms of EM fields. It gives,

For the antisymmetric sector:
\begin{equation}
	\vec{\nabla}\cdot\left[4k_1(a_1 + \beta_1)\vec{E}_A + 4k_2 \left(a_2 +\beta_2\right)\vec{E}_U + 2k_+ \vec{E}_- + 2k_-\vec{E}_+ + 2\left(a_1k_1 + a_2k_2\right)\vec{\mathbf{e}}_+-\right] = \rho^{T}_q
\end{equation}
with
\begin{equation}
	\rho^{T}_q \equiv -q\left\{4a_3\left\{\vec{V}^+\cdot\vec{E}_-\right\}\right\} +\left\{b_3 \vec{V}^+\cdot\left[\vec{\mathbf{e}}_{-A} + \vec{\mathbf{e}}_{-U} \right]\right\}
\end{equation}
and
\begin{multline}
	\vec{\nabla}\times \left[4k_1\left(a_1 + \beta_1\right)\vec{B}_A + 4 \left(a_2 + \beta_2\right)\vec{B}_U + 2k_+a_3 \vec{B}_- + 2k_-a_3 \vec{B}_+ + 2\left(a_1k_1 + a_2k_2\right)\vec{\mathbf{b}}_{+-}\right]+\\
	-\frac{\partial}{\partial t }\left[4k_1\left(a_1 + \beta_1\right)\vec{E}_A + 4k_2 \left(a_2 + \beta_2\right)\vec{E}_U + 2k_+a_3 \vec{E}_- + 2k_-a_3 \vec{E}_-\right] = -\vec{j}^{T}_q
\end{multline}
with
\begin{equation}
	\vec{j}^{T}_{q} \equiv -q\left\{4a_3 Im\left\{\vec{E}_- \cdot \phi^{+} + \vec{B}_- \times \vec{V^+}\right\} + 4b_3 Im\left\{\vec{\mathbf{e}}_{-A} \phi^+ + \vec{\mathbf{b}}_{-A} \times \vec{V}^+\right\}\right\}
\end{equation}

The above equations are expressing the electric charge fundamental law. Showing through a so-called gauge parameter homothety a more factual electric charge behavior in terms of electric and magnetic fields. Diversely from Maxwell, it extends the q meaning. It rewrites the continuity equation, $\frac{\partial \rho^{T}_{q}}{\partial t} + \nabla \cdot \vec{j}^{T}_q = 0$, where the charge density and current, $\rho^{T}_{q}$ and $\vec{j}^{T}_{q}$, are no more expressed in terms of electric charge but in terms of an EM fields flux. Including, eqs. (3.2-3.5) is registering the presence of potential fields connected with granular and collective EM fields through the coupling constant q. Also, chargeless fields $A_{\mu}$ and $U_{\mu}$ are participating in this electric charge dynamics.  

For longitudinal sector:
\begin{multline}
	\frac{\partial}{\partial t}\{4(11k_1\rho_1 + \frac{1}{2}k_2 \rho_2\beta_1 + 11\rho_1\rho_2 + \frac{1}{4}k_2 \xi_{(12)})S^{\alpha}_{\alpha A} +4(11k_2\rho_2 + \frac{1}{2}k_1 \rho_1\beta_2 + \\
	+11\rho_1\rho_2 + \frac{1}{4}k_1 \xi_{(12)})S^{\alpha}_{\alpha U}\} = \rho^{L}_q
\end{multline}
with
\begin{eqnarray}
	&&\rho^{L}_{q} \equiv -q\{4Im\{\beta_+\beta_-\vec{V}^+\vec{S}_- +\left(16\rho_+\rho_- + \rho_+\beta_- +\rho_-\beta_+\right)\phi^+S^{\alpha}_{\alpha -}\} +  \nonumber
	\\
	&&Im\{\beta_+\left(\vec{s}_{-A} + \vec{s}_{-U}\right)\cdot \vec{V}^+ +\left(\beta_+ + 17\rho_+\right)\left(s^{\alpha}_{\alpha -A} + s^{\alpha}_{\alpha -U}\right)\phi^+\}\} 
\end{eqnarray}
and
\begin{multline}
	\partial_i \{4(11k_1\rho_1 + \frac{1}{2}k_2 \rho_2\beta_1 + 11\rho_1\rho_2 + \frac{1}{4}k_2 \xi_{(12)})S^{\alpha}_{\alpha A} +4(11k_2\rho_2 + \frac{1}{2}k_1 \rho_1\beta_2 + \\
	+11\rho_1\rho_2 + \frac{1}{4}k_1 \xi_{(12)})S^{\alpha}_{\alpha U}\} = j_{iq} 
\end{multline}
with
\begin{eqnarray}
	&&j^{L}_{iq} \equiv -q \{4Im\{\beta_+\beta_-V_j^+ S_i^{j-} + \left(16\rho_+\rho_- + \rho_+ \beta_- + \rho_- \beta_+\right)V_i^{+}S^{\alpha}_{\alpha -} \} +\nonumber
	\\
	 &&+4Im\{ \beta_+ \left(s^{j}_{i -A}\right)V^{+}_{j} + s^{j}_{i -U}V^+_j+(17\rho_+ + \beta_+)(s^{\alpha}_{\alpha -A} \nonumber
	 \\
	 &&+ s^{\alpha}_{\alpha -U})V_{i}^{+}\}\}
\end{eqnarray}
introducing another continuity equations $\frac{\partial \vec{j}^{L}_{q}}{\partial t} + \vec{\nabla} \rho^{L}_{q} =0$

A new perspective on electric charge physicality is discovered. The above expressions are showing a fields flow conducting $\{+,0,-\}$ charges. Different electric charge conservation laws at transverse and longitudinal sectors are detected.  Although under the same $q$, they differ dynamically. Showing that, more than a fine structure constant, electric charge is related to a field's flux.

EM is powered by an electric charge. However, Maxwell and QED do not say what electric charge is. No more than classifying in positive and negative charges. On other hand, physics understood from Heisenberg isospin and Gell-Mann-Nishijima strangeness that quantum numbers are deeper than electric charge [31]. Quantum numbers are associated not only with conservation laws as to group symmetry generators. Then, given $U_{q}(1)$, it is expected two quantum numbers are associated with U(1) and SO(2) respectively. It will correspond to, two associated corresponding conservation laws. A justification for the above, two electric charge continuity equations appear associated with spin-1 and spin-0.

\section{Four-Four Maxwell equations}

\indent Electromagnetism is being enlarged to physics, considering four bosons interconnected by electric charge symmetry. However, the corresponding equations are not only derived by Euler-Lagrange and granular Bianchi identities. Differently, from usual electrodynamics [32], constitutive physics is manifested [23]. It includes algebraic identities, Noether identities, and collective Bianchi identities. Together they make a system identified as the Four-Four Maxwell equations.

A constitutive EM is developed beyond the minimal action principle. A systemic behaviour is conducted from the quadruplet $\{A_{\mu}, U_{\mu}, V_{\mu}^{\pm}\}$ electric charge symmetry. While equations appear. They rewrite the Gauss, Amp$\grave{e}$re, Faraday laws with modifications.
\\
\\

For $A^{T}_{\mu}$ (spin-1):
\\
\\

The corresponding whole Gauss law is
\begin{equation}\label{Whole Gauss equation}
	\vec{\nabla} \cdot \left\{4\left(a_1 + \beta_1 + a_1k_1\right)\vec{E}_A + 2 b_1 \left[\vec{\mathbf{e}}_{AU} + \left( 1 + k_1\right)\vec{\mathbf{e}}_{+-}\right]\right\} + l^{T}_{A} + c^{T}_{A} = M^{T}_{IA} + \rho^{T}_{AW} - \rho^{T}_{ q} - k_2 \rho^{T}_{U}
\end{equation}
with
\begin{equation}
	l^{T}_{A} \equiv -4 \mathbf{e}_{[12]}\vec{U} \cdot \vec{\mathbf{e}}_{AU},
\end{equation}
\begin{multline}
	c^{T}_{A} \equiv -4 \mathbf{e}_{[12]} \vec{U} \cdot \vec{\mathbf{e}}_{+-} - \vec{V}^{-} \cdot \left[\sqrt{2}\left(\mathbf{e}_{[13]} - i\mathbf{e}_{[14]}\right)\vec{\mathbf{e}}_{+A} + \sqrt{2}\left(\mathbf{e}_{[13]} - i \mathbf{e}_{[14]}\right)\vec{\mathbf{e}}_{+U}\right]+\\
	- \vec{V}^{+} \cdot \left[\sqrt{2}\left(\mathbf{e}_{[13]} + i\mathbf{e}_{[14]}\right)\vec{\mathbf{e}}_{-A} + \sqrt{2}\left(\mathbf{e}_{[13]} - i \mathbf{e}_{[14]}\right)\vec{\mathbf{e}}_{-U}\right],
\end{multline}
and
\begin{equation}
	\rho^{T}_{AW} \equiv -2 \mathbf{e}_{[12]}\left(\vec{E}_A + \vec{E}_U\right)\cdot \vec{U} - \sqrt{2}\left(\mathbf{e}_{[13]} + i\mathbf{e}_{[14]}\right)\vec{E}_- \cdot \vec{V}^+ \cdot \left[\sqrt{2}\left(\mathbf{e}_{[13]} - i\mathbf{e}_{[14]}\right)\vec{E}_+\cdot\vec{V}^-\right],
\end{equation}
\begin{equation}
	M^{T A}_{I} \equiv 2k_2\mathbf{m}^2_{U}\phi_U + k_{-}\mathbf{m}^2_{V}\phi^+ + k_+\mathbf{m}^2_{V}\phi^-.
\end{equation}

At LHS eq. (4.1) is expressing the EM fields strengths dynamics plus fields massive terms where $l_{AT}$ means the London term associated to the  $U_{\mu}$ field, $c_{AT}$ a conglomerate term grouping different fields. On RHS,  $\rho^{T}_{A}$ corresponds to field densities, and $M_{AT}$ is the mass source associated with fields. Notice that $\mathbf{m}^2_{U}$,$\mathbf{m}^2_{V}$ are just mass parameters introduced without requiring the Higgs mechanism. Electric charge density is expressed by $\rho_{q}^{T}$ at eq. (3.3); $\rho^{T}_{U}$ is in Appendices A

The whole Amp$\grave{e}$re law is
\begin{multline}\label{Whole Ampere-Maxwell for A}
	\vec{\nabla} \times \left[4\left(a_1 + \beta_1 + a_1 k_1\right)\vec{B}_A + 2b_1 \left(\vec{b}_{AU} + \left(1+k_1\right)\vec{b}_{+-}\right)\right]-\frac{\partial}{\partial t}[4(a_1 + \beta_1 + a_1 k_1)\vec{E}_A + \\
	+2 b_1 (\vec{\mathbf{e}}_{AU} + (1 + k_1) \vec{\mathbf{e}}_{+-}) ]+ \vec{l}_{AT} + \vec{c}_{AT} = \vec{M}^{A}_{IT} + \vec{j}^{W}_{AT} - \vec{j}_{q T} - k_2 \vec{j}_{UT}
\end{multline}
with
\begin{equation}
	\vec{l}^{T}_{A} \equiv -4 \mathbf{e}_{[12]}\left\{\vec{\mathbf{e}}_{AU} \phi_U + \vec{b}_{AU} \times \vec{U} \right\},
\end{equation}
and
\begin{eqnarray}
	&&\vec{c}^{T}_{A} \equiv  - \sqrt{2}\left(\mathbf{e}_{[13]} - i\mathbf{e}_{[14]}\right)\left[\left(\vec{\mathbf{e}}_{+A} + \vec{\mathbf{e}}_{+U}\right) \cdot \phi^- + \left(\vec{b}_{+A} + \vec{b}_{-U}\right)\times \vec{V}^-  \right] + \nonumber
	\\
	&& - \sqrt{2}\left(\mathbf{e}_{[13]} + i\mathbf{e}_{[14]}\right)\left[\left(\vec{\mathbf{e}}_{-A}+ \vec{\mathbf{e}}_{-U}\right)\cdot \phi^+ + \left(\vec{b}_{-A} + \vec{b}_{+U}\right)\times \vec{V}^+\right]+  
	\\
	&&-4 \mathbf{e}_{[12]}\left\{\vec{\mathbf{e}}_{+-} \cdot \phi_U + \vec{b}_{+-} \times \vec{U}\right\},\nonumber
\end{eqnarray}
\begin{eqnarray}
	&&\vec{j}^{T}_{AW} \equiv -2\mathbf{e}_{[12]}\left[\left(\vec{E}_{A} + \vec{E}_{U}\right)\cdot \phi_U + \left(\vec{B}_A + \vec{B}_{U}\right)\times \vec{U}\right]\nonumber
	\\
	 &&-\sqrt{2}\left(\mathbf{e}_{[13]} - i\mathbf{e}_{[14]}\right)\left[\vec{E}_{+}\cdot \phi^- + \left(\vec{B}_{+} \times \vec{V}^{-}\right)\right]+
	\\
	&&-\sqrt{2}\left(\mathbf{e}_{[13]} + i\mathbf{e}_{[14]}\right)\left[\vec{E}_{-}\cdot \phi^+ + \vec{B}_{-}\times \vec{V}^{+}\right]\nonumber,
\end{eqnarray}
\begin{equation}
	\vec{M}^{T}_{IA} \equiv 2k_2\mathbf{m}^2_{U} \vec{U} + k_{-}\mathbf{m}^2_{V}\vec{V}^{+} + k_+\mathbf{m}^2_{V}\vec{V}^{-}.
\end{equation}
Electric charge current $\vec{j}_{q T}$ is written at eq. (3.5).

It yields the following constitutive conservation law.
\begin{equation}\label{Lei de conservação W}
	\frac{\partial\rho^{T}_{CA}}{\partial t} + \nabla \cdot \vec{j}^{T}_{CA} =0
\end{equation}
where
\begin{equation}
	\rho^{T}_{CA} = -l^{T}_{A} - c^{T}_{A} + M^{T}_{IA} + \rho^{T}_{AW} - k_2 \rho^{T}_{U}
\end{equation}
and
\begin{equation}
	\vec{j}^{T}_{CA} = -\vec{l}^{T}_{A} - \vec{c}^{T}_{A} + \vec{M}^{T}_{IA} + \vec{j}^{T}_{AW} - k_2 \vec{j}^{T}_{U}
\end{equation}

A mass continuity equation is rewritten from eq. (4.11). It gives,

\begin{equation}
	\frac{\partial M^{T}_{IA}}{\partial t} + \nabla\cdot\vec{M}^{T}_{IA} = l^{T}_{A} + c^{T}_{A} - \rho^{T}_{AW} + k_{2}\rho^{T}_{U}
\end{equation}

A photon constitutive fundamental equation is derived. As radiation, without charge and mass, the photon behaviour is extended as a field working as its own source. It incorporates photon dynamics with nonlinear granular and collective fields strengths. New features appear. A photon generates its own EM field, London, and conglomerate fields as masses terms, fields currents, and massive sources. Also, physical entities such as electric charge and mass with a physicality given by continuity equations depending on fields.

Eqs. (4.1-4.14) are introducing a shift to Gauss law and Maxwell-Amp$\grave{e}$re law. Actually, three reasons are offered to modify the Maxwell-Amp$\grave{e}$re law. They are by introducing mass according to de Broglie-Proca, Lorentz symmetry violation, and effective nonlinear EM [33]. An analysis is being studied under the MMS satellite date [34]. Differently, the above equation modifications are due to a fundamental EM.

For $U^{T}_{\mu}$: the massive photon constitutive Gauss law is
\begin{eqnarray}\label{Whole Gauss equation for U}
	&&\vec{\nabla} \cdot \left\{4\left(a_2 + \beta_2 + a_2 k_2\right)\vec{E}_U + 2b_2\left[\vec{\mathbf{e}}_{AU} + \left(1 + k_1\right)\vec{\mathbf{e}}_{+-}\right]\right\}\nonumber
	\\
	&&-2\mathbf{m}^2_{U}\phi_U + l_{UT} + c_{UT} = M^{U}_{IT} + \rho^{W}_{UT} - k_1\rho_{AT} - \rho_{NT}
\end{eqnarray}
with
\begin{equation}
	l^{T}_{U } \equiv -4 \mathbf{e}_{[12]} \vec{A}\cdot \vec{\mathbf{e}}_{AU},
\end{equation}
\begin{multline}
	c^{T}_{U} \equiv -4\mathbf{e}_{[12]}\vec{A}\cdot \vec{\mathbf{e}}_{+-}-\vec{V}^{-}\cdot\left[\sqrt{2}\left(\mathbf{e}_{[23]} - i\mathbf{e}_{[14]}\right)\vec{\mathbf{e}}_{+A} + \sqrt{2}\left(\mathbf{e}_{[23]}-i\mathbf{e}_{[24]}\right)\vec{\mathbf{e}}_{+U}\right]+\\
	-\vec{V}^{+}\cdot\left[\sqrt{2}\left(\mathbf{e}_{[23]} + i\mathbf{e}_{[14]}\right)\vec{\mathbf{e}}_{-A} + \sqrt{2}\left(\mathbf{e}_{[23]}-i\mathbf{e}_{[24]}\right)\vec{\mathbf{e}}_{-U}\right],
\end{multline}
and
\begin{multline}
	\rho^{T}_{U W} \equiv +2i \mathbf{e}_{[12]} \left(\vec{E}_{A} + \vec{E}_{U}\right)\cdot \vec{A} - \sqrt{2}\left(\mathbf{e}_{[23]} - i\mathbf{e}_{[24]}\right)\vec{E}_{+} \cdot \vec{V}^{-} - \sqrt{2}\left(\mathbf{e}_{[23]} +i \mathbf{e}_{[24]}\right)\vec{E}_- \cdot \vec{V}^{+} 
\end{multline}
\begin{equation}
	M^{T}_{IU} \equiv k_- \mathbf{m}^{2}_{V} \phi^{+} + k_+\mathbf{m}^{2}_{V} \phi^{-}.
\end{equation}

The corresponding constitutive Amp$\grave{e}$re law is
\begin{eqnarray}\label{Whole Ampere Maxwell for U}
	&&\vec{\nabla}\times \left[4\left(a_2 + \beta_2 a_2 k_2 \right) \vec{B}_{U} + 2b_2\left(\vec{b}_{AU} + \left(1 + k_2\right)\vec{b}_{+-}\right)\right] - \frac{\partial}{\partial t}[4(a_2 + \beta_2 + k_2 a_2)\vec{E}_U +\nonumber 
	\\
	&&+2b_2 \vec{\mathbf{e}}_{AU} + (1+k_2)\vec{\mathbf{e}}_{+-}] + \vec{l}_{UT} + \vec{c}_{UT} = \vec{M}^{U}_{IT} + \vec{j}_{UT} - \vec{j}^{W}_{NT} - k_1 \vec{j}_{AT}
\end{eqnarray}
with
\begin{equation}
	\vec{l}^{T}_{U} \equiv -4\mathbf{e}_{[12]}\left[\mathbf{e}_{AU} \cdot \phi^{-} + \vec{b}_{AU} \times \vec{A}\right],
\end{equation}
\begin{eqnarray}
	&&\vec{c}^{T}_{U} \equiv  - \sqrt{2}\left(\mathbf{e}_{[23]} - i \mathbf{e}_{[24]}\right)\left[\left(\vec{\mathbf{{e}}}_{+A} + \vec{\mathbf{e}}_{+U}\right)\cdot \phi^{-} + \left(\vec{b}_{+A} \vec{b}_{+U}\right)\times \vec{V}^{-}\right]+\nonumber
	\\
	&&-\sqrt{2}\left(\mathbf{e}_{[23]} + i \mathbf{e}_{[24]}\right)\left[\left(\vec{\mathbf{e}}_{-A} + \vec{\mathbf{e}}_{-U}\right)\cdot \phi^{+} + \left(\vec{b}_{-A} + \vec{b}_{-U}\right)\times \vec{V}^{+}\right]+
	\\
	&&-4 \mathbf{e}_{[12]}\left[\vec{\mathbf{e}}_{+-} \cdot \phi_A + \vec{b}_{+-} \times \vec{A}\right]\nonumber
\end{eqnarray}
and
\begin{eqnarray}
	&&\vec{j}^{T}_{U W} \equiv -2 \mathbf{e}_{[12]}\left[\left(\vec{E}_{A} + \vec{E}_{U}\right)\cdot \phi_{A} + \left(\vec{B}_A + \vec{B}_B\right) \times \vec{A}\right] +\nonumber
	\\
	 &&+\sqrt{2}\left(\mathbf{e}_{[23]} - i \mathbf{e}_{[24]}\right)\left[\vec{E}_+ \cdot \phi_{-} + \left(\vec{B}_{+} \times \vec{V}^{-}\right)\right] + 
	 \\
	&&-\sqrt{2}\left(\mathbf{e}_{[23]} + i \mathbf{e}_{[24]}\right)\left[\vec{E}_- \cdot \phi_+ + \left(\vec{B}_{-} \times \vec{V}^{+}\right)\right],\nonumber
\end{eqnarray}
\begin{equation}
	\vec{M}^{U}_{UT} \equiv  k_{-} \mathbf{m}^{2}_{V} \vec{V}^{+} + k_+ \mathbf{m}^{2}_{V} \vec{V}^{-}.
\end{equation}
For $V_{\mu}^{T-}$:

The correspondent constitutive Gauss and Amp$\grave{e}$re laws for a negative charged massive photon are

\begin{equation}\label{Whole Gauss equation V+}
	\vec{\nabla}\cdot \left[2\left(a_3 + \beta_+\beta_-\right)\vec{E}^{+} + 2b_3\left(\vec{\mathbf{e}}_{+A} + \vec{\mathbf{e}}_{+U}\right)\right]-\mathbf{m}_{V}^{2}\phi^{+} +l^{+}_{VT} + c^{+}_{VT} = \rho^{W+}_{VT}.
\end{equation}
with
\begin{equation}
	l^{T+}_{V} = 0,
\end{equation}
\begin{eqnarray}
	&&c^{T+}_{V} \equiv -\sqrt{2}\left(\mathbf{e}_{[13]} - i\mathbf{e}_{[14]}\right)\left[\vec{A}\cdot \left(\vec{\mathbf{e}}_{+A} + \vec{\mathbf{e}}_{+U}\right)\right] 
	\\
	&&-\sqrt{2}\left(\mathbf{e}_{[23]} - i\mathbf{e}_{[24]}\right)\left[\vec{U}\cdot \left(\vec{\mathbf{e}}_{+A} + \vec{\mathbf{e}}_{+U}\right)\right]+
	\\
	&&-4i\mathbf{e}_{[34]}\left[\vec{V}^{+}\cdot \left(\vec{\mathbf{e}}_{+-} + \vec{\mathbf{e}}_{AU}\right)\right],
\end{eqnarray}
\begin{equation}
	\rho^{T+}_{V W} \equiv 2i\mathbf{e}_{[34]}\left(\vec{E}_{A} + \vec{E}_{U}\right)\cdot \vec{V}^{+} - \sqrt{2}\left(\mathbf{e}_{[13]} - i\mathbf{e}_{[14]}\right)\vec{E}_{A} \cdot \vec{V}^{+} - \sqrt{2}\left(\mathbf{e}_{[23]} -i\mathbf{e}_{[24]}\right)\vec{E}_{+} \cdot \vec{U},
\end{equation}
and
\begin{multline}\label{Whole Ampere-Maxwell V+}
	\vec{\nabla} \times \left[2\left(a_3 + \beta_+\beta_-\right)\vec{B}^+ + 2b_3\left(\vec{b}_{+A} + \vec{b}_{+U}\right)\right] - \frac{\partial}{\partial t}[2(a_3 + \beta_+\beta_-)\vec{E}_+ +2b_3 (\vec{\mathbf{e}}_{+A} + \vec{\mathbf{e}}_{+U})] +\\
	\vec{l}^{+}_{TV} + \vec{c}^{+}_{TV} - \mathbf{m}^{2}_{V} \vec{V}^{+} = \vec{j}^{W+}_{VT}
\end{multline}
with
\begin{equation}
	\vec{l}^{T +}_{V} = 0
\end{equation}
\begin{multline}
	\vec{c}^{T+}_{V} \equiv -\sqrt{2}\left(\mathbf{e}_{[13]} - i\mathbf{e}_{[14]}\right)\left[\mathbf{e}_{+A}\cdot \phi_{A} + \vec{b}_{+A}\times \vec{A} + \vec{\mathbf{e}}_{+U} \cdot \phi_{A} + \vec{b}_{+U} \times \vec{A}\right]+\\
	-\sqrt{2}\left(\mathbf{e}_{[23]} - i\mathbf{e}_{[24]}\right)\left[\vec{\mathbf{e}}_{+U}\cdot \phi_U + \vec{b}_{+A} \times \vec{U} + \vec{\mathbf{e}}_{+U} \cdot \phi_U + \vec{b}_{+U} \times \vec{U}\right] + \\
	-4i\mathbf{e}_{[34]}\left[\vec{\mathbf{e}}_{+-} \cdot \phi^+ + \vec{b}_{+-} \times \vec{V}^{+} + \vec{\mathbf{e}}_{AU} \cdot \phi^+  + \vec{b}_{AU} \times \vec{V}^+ \right]
\end{multline}
\begin{eqnarray}
	&&\vec{j}^{T+}_{VW} \equiv -2i\mathbf{e}_{[34]}\left[\left(\vec{E}_{A} + \vec{E}_{U}\right) \cdot \phi^+ + \left(\vec{B}_{A} + \vec{B}_{U}\right)\times \vec{V}^{+}\right]\nonumber
	\\
	&&-\sqrt{2}\left(\mathbf{e}_{[13]} - i\mathbf{e}_{[14]}\right)\left[\vec{E}_+ \cdot \phi_{A} + \vec{B}_+ \times \vec{A}\right]
	\\
	&&-\sqrt{2}\left(\mathbf{e}_{[23]} - i\mathbf{e}_{[24]}\right)\left[\vec{E}_+\cdot \phi_{U} + \vec{B}_+ \times \vec{U}\right]\nonumber
\end{eqnarray}
For $V^{T+}_{\mu }$: similarly, one gets
\begin{equation}\label{Whole Gauss equation V-}
	\vec{\nabla}\cdot \left[2\left(a_3 + \beta_+\beta_-\right)\vec{E}^{-} + 2b_3\left(\vec{\mathbf{e}}_{-A} + \vec{\mathbf{e}}_{-U}\right)\right]-\mathbf{m}_{V}^{2}\phi^{-} +l^{-}_{VT} + c^{-}_{VT} = \rho^{W-}_{VT}
\end{equation}
with
\begin{equation}
	l^{T-}_{V} = 0
\end{equation}
\begin{eqnarray}
	&&c^{T-}_{V} \equiv -\sqrt{2}\left(\mathbf{e}_{[13]} + i\mathbf{e}_{[14]}\right)\left[\vec{A}\cdot \left(\vec{\mathbf{e}}_{-A} + \vec{\mathbf{e}}_{-U}\right)\right]\nonumber 
	\\
	&&-\sqrt{2}\left(\mathbf{e}_{[23]} + i\mathbf{e}_{[24]}\right)\left[\vec{U}\cdot \left(\vec{\mathbf{e}}_{-A} + \vec{\mathbf{e}}_{-U}\right)\right]
	\\
	&&-4i\mathbf{e}_{[34]}\left[\vec{V}^{-}\cdot \left(\vec{\mathbf{e}}_{+-} + \vec{\mathbf{e}}_{AU}\right)\right]\nonumber
\end{eqnarray}
\begin{equation}
	\rho^{T-}_{Vw} \equiv 2i\mathbf{e}_{[34]}\left(\vec{E}_{A} + \vec{E}_{U}\right)\cdot \vec{V}^{-} - \sqrt{2}\left(\mathbf{e}_{[13]} + i\mathbf{e}_{[14]}\right)\vec{E}_{-} \cdot \vec{A} - \sqrt{2}\left(\mathbf{e}_{[23]} + i\mathbf{e}_{[24]}\right)\vec{E}_- \cdot \vec{U}
\end{equation}
and
\begin{eqnarray}\label{Whole Ampere-Maxwell for V-}
	&&\vec{\nabla} \times \left[2\left(a_3 + \beta_+\beta_-\right)\vec{B}^+ + 2b_3\left(\vec{b}_{+A} + \vec{b}_{+U}\right)\right] - \frac{\partial}{\partial t}[2(a_3 + \beta_+\beta_-)\vec{E}_+ \nonumber
	\\
	&&+2b_3 (\vec{\mathbf{e}}_{+A} + \vec{\mathbf{e}}_{+U})] +\vec{l}^{+}_{TV} + \vec{c}^{+}_{TV} - \mathbf{m}^{2}_{V} \vec{V}^{+} = \vec{j}^{W+}_{VT}
\end{eqnarray}
with
\begin{eqnarray}
	&&\vec{c}^{T-}_{V} \equiv -\sqrt{2}\left(\mathbf{e}_{[13]} + i\mathbf{e}_{[14]}\right)\left[\mathbf{e}_{-A}\cdot \phi_{A} + \vec{b}_{-A}\times \vec{A} + \vec{\mathbf{e}}_{-U} \cdot \phi_{A} + \vec{b}_{-U} \times \vec{A}\right]+\nonumber
	\\
	&&-\sqrt{2}\left(\mathbf{e}_{[23]} + i\mathbf{e}_{[24]}\right)\left[\vec{\mathbf{e}}_{-U}\cdot \phi_U + \vec{b}_{-A} \times \vec{U} + \vec{\mathbf{e}}_{-U} \cdot \phi_U + \vec{b}_{-U} \times \vec{U}\right] + 
	\\
	&&-4i\mathbf{e}_{[34]}\left[\vec{\mathbf{e}}_{+-} \cdot \phi^- + \vec{b}_{+-} \times \vec{V}^{-} + \vec{\mathbf{e}}_{AU} \cdot \phi^-  + \vec{b}_{AU} \times \vec{V}^-\nonumber \right]
\end{eqnarray}
\begin{eqnarray}
	&&\vec{j}^{T-}_{VW} \equiv -2i\mathbf{e}_{[34]}\left[\left(\vec{E}_{A} + \vec{E}_{U}\right) \cdot \phi^- + \left(\vec{B}_{A} + \vec{B}_{U}\right)\times \vec{V}^{-}\right]\nonumber
	\\
	&&-\sqrt{2}\left(\mathbf{e}_{[13]} + i\mathbf{e}_{[14]}\right)\left[\vec{E}_- \cdot \phi_{A} + \vec{B}_- \times \vec{A}\right]
	\\
	&&-\sqrt{2}\left(\mathbf{e}_{[23]} + i\mathbf{e}_{[24]}\right)\left[\vec{E}_-\cdot \phi_{U} + \vec{B}_- \times \vec{U}\right]\nonumber
\end{eqnarray}
Eqs. (4.14-4.17) follow the same conservation law as eq. (4.11)

Lorentz's symmetry relates to the presence of the physical spin-0 sector. It yields also two kinds of equations and corresponding conservation laws.

For $A^{L}_{\mu}$ (spin-0):

The time-dependent equation is
\begin{eqnarray}
	&&\partial^{0}\left[s_{11}S^{\alpha}_{\alpha A} + c_{11}\left(s^{\alpha}_{\alpha AA} + s^{\alpha}_{\alpha UU}+ s^{\alpha}_{\alpha AU} + s^{\alpha}_{\alpha +-}\right)\right] + l^{0}_{AL} + c^{0}_{AL} = j^{W 0}_{AL} - J^{0}_{qL} \nonumber
	\\
	&&+ t_{11}\mathbf{m}U^{0}_{UL} + \frac{1}{2}t_{11}J^{0}_{UL}
\end{eqnarray}
with
\begin{multline}
	l^{0}_{AL} = -\mathbf{e}_{(11)}\left(84 A^0 s^{\alpha}_{\alpha AA} + 4A^{0}s^{\alpha}_{\alpha AA} + 4s^{0i}_{+-}A_{i}\right) + 4\mathbf{e}_{(12)}\left(s^{00}_{AA}A_0 + s^{0i}_{AA}A_{i}\right)\\
	-\mathbf{e}_{(12)}\left(68s^{\alpha}_{\alpha AA} U_{0} + 80 s^{\alpha}_{\alpha UU}U_0 + s^{\alpha}_{\alpha AU}U_0 \right),
\end{multline}

\begin{eqnarray}
	&&c^{0}_{AL} = -\mathbf{e}_{(11)}\left(4s^{00}_{+-}A_{0} + 4 s^{0i}_{+-}A_{i} + 68s^{\alpha}_{\alpha +-} A_{0}\right)-72\mathbf{e}_{(12)}s^{\alpha}_{\alpha +-}U_0 
	\\
	&&-\sqrt{2}\left(\mathbf{e}_{(13)}-i\mathbf{e}_{(14)}\right)\left(s^{00}_{+A}V^{-}_{0} + s^{0i}_{+A}V_{i}^{-} + 18s^{\alpha}_{\alpha+A}V_{0}^{-} + s^{0i}_{+U}V_{i}^{-} + 18s^{\alpha}_{\alpha +U} V_{0}^{-}\right)
	\\
	&&-\sqrt{2}\left(\mathbf{e}_{(13)}+i\mathbf{e}_{(14)}\right)\left(s^{00}_{-A}V^{+}_{0} + s^{0i}_{-A}V_{i}^{+} + 18s^{\alpha}_{\alpha-A}V_{0}^{+} + s^{0i}_{-U}V_{i}^{+} + 18s^{\alpha}_{\alpha -U} V_{0}^{+}\right) 
\end{eqnarray}
and
\begin{eqnarray}
	&&j^{0L}_{A} = 2 \left(\mathbf{e}_{(11)} + \mathbf{e}_{(12)}\right) \big\{ \left(\beta_1\vec{S}_A + \beta_2 \vec{S}_U \right) \cdot \left(\vec{A} + \vec{U}\right)  \nonumber
	\\
	&&+ \left[\left( \beta_1 + 17\rho_1 \right)S^{\alpha}_{\alpha A} + \left(\beta_2 +17\rho_2\right) S^{\alpha}_{\alpha U} \right] \left(\phi_A + \phi_U\right) \big\}\nonumber
	\\
	&&+\sqrt{2}\left(\mathbf{e}_{(13)} - i\mathbf{e}_{(14)}\right)\left[\beta_+ \vec{S}^+\vec{V}^- + \left(\beta_+ + 17\rho_+\right)S^{\alpha}_{\alpha +}\phi^-\right] \nonumber
	\\
	&&- \beta_1\mathbf{e}_{(11)}\left(S^{\alpha}_{\alpha A} \phi_A + 2 \vec{S}_A \cdot \vec{A}\right)+
	\\
	&&-\beta_1\mathbf{e}\left(S^{\alpha}_{\alpha A}\phi_U + S^{\alpha}_{\alpha U}\phi_A + \vec{S}_A \cdot \vec{U} + \vec{S}_U \cdot \vec{A}\right) \nonumber
	\\
	&&- \beta_1\left(\mathbf{e}_{(33)} + \mathbf{e}_{(44)}\right) \left(S^{\alpha}_{\alpha +}\phi^- + S^{\alpha}_{\alpha -}\phi^+ + \vec{S}_+\vec{V}^- + \vec{S}_-\vec{V}^+\right)\nonumber
\end{eqnarray}

The correspondent vectorial equation is
\begin{eqnarray}
	&&\partial^{i}\left[s_{11}S^{\alpha}_{\alpha A} + c_{11}\left(s^{\alpha}_{\alpha AA} + s^{\alpha}_{\alpha AU} + s^{\alpha}_{\alpha UU} + s^{\alpha}_{\alpha+-}\right)\right]+l^{i}_{AL} +\nonumber
	\\
	&&+ c^{i}_{AL} = - j^{i}_{AL} +J^{i}_{qL} - t_{11}\mathbf{m}_{U}^2U^{i}-\frac{1}{2}t_{11}j^{i}_{UL}
\end{eqnarray}
with
\begin{multline}
	l^{i}_{AL} = -\mathbf{e}_{(11)}\left[4s^{i0}_{+-}A_0 + 4s^{ij}_{+-}A_{j} + 84s^{\alpha}_{\alpha UU}A^{i}\right]-\mathbf{e}_{(12)}\left(s^{0i}_{AA}A_{0} + s^{ij}_{AA}A_{j}\right)\\
	-\mathbf{e}_{(12)}\left(68s^{\alpha}_{\alpha AA}U^{i} + 8s^{0i}U_{0}+8s^{ij}_{UU}U_{j}+76s^{\alpha}_{\alpha AU}U^{i}\right),
\end{multline}
\begin{multline}
	c^{i}_{AL} = -\mathbf{e}_{(11)}\left(s_{+-}^{i0}A_{0} + s^{ij}_{+-}A_{j} + 68s^{\alpha}_{\alpha +-}A^{i}\right) - 72s^{\alpha}_{\alpha +-}A^{i}\\
	-\sqrt{2}\left(\mathbf{e}_{(13)} - i\mathbf{e}_{(14)}\right)\left[s^{i0}_{+A}V_{0}^{-}+s^{ij}_{+A}V_{j}^{-} + 18s^{\alpha}_{\alpha +A}V^{i-}+s^{i0}_{+U}V^{-}_{0} +s^{ij}_{+U}V^{-}_{j} + 18s^{\alpha}_{\alpha +U}V^{i-}\right]+\\
	-\sqrt{2}\left(\mathbf{e}_{(13)} + i\mathbf{e}_{(14)}\right)\left[s^{i0}_{-A}V_{0}^{+}+s^{ij}_{-A}V_{j}^{+} + 18s^{\alpha}_{\alpha -A}V^{i+}+s^{i0}_{-U}V^{+}_{0} +s^{ij}_{-U}V^{+}_{j} + 18s^{\alpha}_{\alpha -U}V^{i+}\right]
\end{multline}
and
\begin{multline}
	j^{i}_{AL} = 2\beta_1\left\{S^{i0}_A\left(\mathbf{e}_{(11)}A_0 + \mathbf{e}_{(12)}U_0\right) + S^{ij}_A\left( \mathbf{e}_{(11)}A_j + \mathbf{e}_{(12)}U_j \right) + S^{\alpha}_{\alpha A}\left( \mathbf{e}_{(11)}A^i + \mathbf{e}_{(12)}U^i \right)\right\}\\
	+2\beta_2\left\{S^{i0}_U\left(\mathbf{e}_{(11)}A_0 + \mathbf{e}_{(12)}U_0\right) + S^{ij}_U\left( \mathbf{e}_{(11)}A_j + \mathbf{e}_{(12)}U_j \right) + S^{\alpha}_{\alpha U}\left( \mathbf{e}_{(11)}A^i + \mathbf{e}_{(12)}U^i \right)\right\}\\
	34\rho_1S^{\alpha}_{\alpha A}\left(\mathbf{e}_{(11)}A^i + \mathbf{e}_{(12)}U^i\right) + 34\rho_2S^{\alpha}_{\alpha U}\left(\mathbf{e}_{(11)}A^i + \mathbf{e}_{(12)}U^i\right) + \sqrt{2}\left(\mathbf{e}_{(13)} - i\mathbf{e}_{(14)}\right)[\beta_+S^{i0}_+ V_0^- +\\ 
	+\beta_+S^{ij}_+V_j^- \left(\beta_+ + 17\rho_+\right)S^{\alpha}_{\alpha +}V^{i-}] + \sqrt{2}\left(\mathbf{e}_{(13)} + i\mathbf{e}_{(14)}\right)[\beta_-S^{i0}_- V_0^+ +\beta_-S^{ij}_-V_j^+ \left(\beta_- + 17\rho_-\right)S^{\alpha}_{\alpha -}V^{i+}]
\end{multline}

A realistic four scalar photon's quadruplet appears. However, the scalar photon physics contains peculiarities.These scalar field's strength dynamics can be removed by the gauge fixing term. A pure photonic is expressed through the London, conglomerate, and current terms.

The following conservation law is obtained

\begin{equation}
	\partial^{i}\rho_{L}^{A C} + \partial_{0}j^{i L}_{A C}=0
\end{equation}
where
\begin{equation}
	\rho^{L}_{A C} = - l^{0}_{AL} - c^{0}_{AL} + j^{W 0}_{AL} + t_{11}\mathbf{m}U^{0}_{UL} + \frac{1}{2}t_{11}J^{0}_{UL}
\end{equation}
\begin{equation}
	j^{i L}_{A C} = -l^{i}_{AL} - c^{i}_{AL} - j^{i}_{AL} - t_{11}\mathbf{m}_{U}^2U^{i}-\frac{1}{2}t_{11}j^{i}_{UL}
\end{equation}

For $U^{L}_{\mu}$:

The spin-0 massive photon is associated to the following temporal and spatial equations,
\begin{eqnarray}
	&&\partial_{0}\left[s_{22}S^{\alpha}_{\alpha U} + c_{22}\left(s^{\alpha}_{\alpha AA} + s^{\alpha}_{\alpha UU} + s^{\alpha}_{\alpha AU} + s^{\alpha}_{\alpha +-}\right)\right] - 2\mathbf{m}^{2}_{U}U^{0} + l_{UL}^{0}
	\\
	&&+c^{0}_{UL} = j^{0}_{UL} - k_{1}j^{W 0}_{AL} - J^{0}_{qL}
\end{eqnarray}
with
\begin{multline}
	l^{0}_{UL} = -\mathbf{e}_{(12)}\left[s^{00}_{UU}A_0 + 4s^{0i}_{UU}A_{i} + 8s^{00}_{AU}A_{0} + 8s^{0i}_{AU}A_{i} + 72s^{\alpha}_{\alpha AA}A^{0} + 4s^{\alpha}_{\alpha}A^{0} + 68s^{\alpha}_{\alpha UU}A^{0}\right] - \\
	-\mathbf{e}_{(22)}\left[s^{00}_{UU}U_0 + 4s^{0i}_{UU}U_{i} + 4s^{\alpha}_{\alpha UU}U^{0} + 4s^{00}_{AA}U_{0} + 4s^{0i}U_{i} + s^{\alpha}_{\alpha}U^{0} + 72s^{\alpha}_{\alpha AU}U^{0}\right]+\\
	+4\mathbf{e}_{(12)}\left[s^{00}_{AU}U_0 + s^{0i}_{AU}U_{i}\right],
\end{multline}
\begin{eqnarray}
	&&c^{0}_{UL} = -72\mathbf{e}_{(12)}s^{\alpha}_{\alpha+-}A^{0}-\mathbf{e}_{(12)}s^{\alpha}_{\alpha +-}U^{0}-\mathbf{e}_{(22)}\left[4s^{00}_{+-}U_0 + 4s^{0i}_{+-}U_{i} + 68s^{\alpha}_{\alpha +-}U^{0} \right]+\nonumber
	\\
	&&-\sqrt{2}\left(\mathbf{e}_{(23)} -i\mathbf{e}_{(24)}\right)[s^{0i}_{+A}V^{-}_{i}+s^{00}_{+A}V^{-}_{i}+18s^{\alpha}_{\alpha +A}V^{0-} + s^{00}_{+U}V^{-}_{0} + s^{0i}_{+U}V^{-}_{i} \nonumber
	\\
	&&+ 18s^{\alpha}_{\alpha +U}V^{0-}]-\sqrt{2}\left(\mathbf{e}_{(23)} +i\mathbf{e}_{(24)}\right)[s^{0i}_{-A}V^{+}_{i}+s^{00}_{-A}V^{+}_{i}+18s^{\alpha}_{\alpha -A}V^{0+} + s^{00}_{-U}V^{+}_{0} \nonumber
	\\
	&&+ s^{0i}_{-U}V^{+}_{i} + 18s^{\alpha}_{\alpha -U}V^{0+}]
\end{eqnarray}
and
\begin{eqnarray}
	&&j^{0L}_{U} = 2 \left(\mathbf{e}_{(22)} + \mathbf{e}_{(12)}\right) \big\{ \left(\beta_1\vec{S}_A + \beta_2 \vec{S}_U \right) \cdot \left(\vec{A} + \vec{U}\right)  \nonumber
	\\
	&&+ \left[\left( \beta_1 + 17\rho_1 \right)S^{\alpha}_{\alpha A} + \left(\beta_2 +17\rho_2\right) S^{\alpha}_{\alpha U} \right] \left(\phi_A + \phi_U\right) \big\}\nonumber
	\\
	&&+\sqrt{2}\left(\mathbf{e}_{(23)} - i\mathbf{e}_{(24)}\right)\left[\beta_+ \vec{S}^+\vec{V}^- + \left(\beta_+ + 17\rho_+\right)S^{\alpha}_{\alpha +}\phi^-\right] 
	\\
	&&- \beta_1\mathbf{e}_{(22)}\left(S^{\alpha}_{\alpha U} \phi_U + 2 \vec{S}_U \cdot \vec{U}\right)-\beta_2\mathbf{e}_{(12)}\big(S^{\alpha}_{\alpha A}\phi_U +\nonumber
	\\
	&&S^{\alpha}_{\alpha U}\phi_A + \vec{S}_A \cdot \vec{U} + \vec{S}_U \cdot \vec{A}\big) - \beta_2\left(\mathbf{e}_{(33)} + \mathbf{e}_{(44)}\right) \big(S^{\alpha}_{\alpha +}\phi^-\nonumber
	\\
	 &&+ S^{\alpha}_{\alpha -}\phi^+ + \vec{S}_+\vec{V}^- + \vec{S}_-\vec{V}^+\big)\nonumber
\end{eqnarray}

The Amp$\grave{e}$re longitudinal massive photon is
\begin{eqnarray}
	&&\partial^{i}\left[s_{22}S^{\alpha}_{\alpha U} +c_{22}\left(s^{\alpha}_{\alpha AA} 6 s^{\alpha}_{\alpha UU} + s^{\alpha}_{\alpha AU} + s^{\alpha}_{\alpha +-}\right)\right] -2\mathbf{m}^{2}_{U}U^{i}_{L}+l^{i}_{UL} \nonumber
	\\
	&&+ c^{i}_{UL} = j^{i}_{UL} -k_{1}j^{i}_{AL}-j^{i}_{qL}
\end{eqnarray}
with
\begin{multline}
	l^{i}_{UL} = -\mathbf{e}_{(12)}\left[4s^{0i}_{UU}A_{0}+4s^{ij}A_{j} + 8s^{0i}_{AU}A_{0} + 8s^{ij}_{AU}A_{j} + 72s^{\alpha}_{\alpha AA}A^{i} + 4s^{\alpha}_{\alpha AU}A^{i}+68s^{\alpha}_{\alpha UU}A^{i}\right] - \\
	\mathbf{e}_{(22)}\left[4s^{0i}_{UU}U_{0} + 4s^{ij}_{UU}U_{j} + 8s^{\alpha}_{\alpha UU}U^{i} + 4s^{0i}_{AA}U_0 + 4s^{ij}_{AA}U_{j} + 4s^{\alpha}_{\alpha AA}U^{i} +72s^{\alpha}_{\alpha AU}U^{i}\right]-\\
	+4\mathbf{e}_{(12)}\left[s^{0i}_{AU}U_0 + s^{ij}_{AU}U_j\right],
\end{multline}
\begin{multline}
	c^{i}_{UL} =72\mathbf{e}_{(12)}s^{\alpha}_{\alpha +-}A^{i}-\mathbf{e}_{(22)}\left[4s^{\alpha}_{\alpha +-}U^{i}-4s^{0i}U_0 +4s^{ij}_{+-}+68s^{\alpha}_{\alpha +-}U^{i}\right]+\\
	-\sqrt{2}\left(\mathbf{e}_{(23)}-\mathbf{e}_{(24)}\right)\left[s^{0i}_{+A}V^{-}_{0}+s^{ij}_{+A}V^{-}_{j}+18s^{\alpha}_{\alpha +A}V^{-i}+s^{i0}_{+U}V_{0}^{-}+s^{ij}_{+U}V^{-}_{j}+s^{\alpha}_{\alpha +U}V^{-i}\right]\\
	-\sqrt{2}\left(\mathbf{e}_{(23)}+\mathbf{e}_{(24)}\right)\left[s^{0i}_{-A}V^{+}_{0}+s^{ij}_{-A}V^{+}_{j}+18s^{\alpha}_{\alpha -A}V^{+i}+s^{i0}_{-U}V_{0}^{+}+s^{ij}_{-U}V^{+}_{j}+s^{\alpha}_{\alpha -U}V^{+i}\right],
\end{multline}
and
\begin{multline}
	j^{i}_{UL} = 2\beta_2\left\{S^{i0}_U\left(\mathbf{e}_{(22)}U_0 + \mathbf{e}_{(12)}A_0\right) + S^{ij}_U\left( \mathbf{e}_{(22)}U_j + \mathbf{e}_{(12)}A_j \right) + S^{\alpha}_{\alpha U}\left( \mathbf{e}_{(22)}U^i + \mathbf{e}_{(12)}A^i \right)\right\}\\
	+2\beta_1\left\{S^{i0}_A\left(\mathbf{e}_{(22)}U_0 + \mathbf{e}_{(12)}A_0\right) + S^{ij}_A\left( \mathbf{e}_{(22)}U_j + \mathbf{e}_{(12)}A_j \right) + S^{\alpha}_{\alpha A}\left( \mathbf{e}_{(22)}U^i + \mathbf{e}_{(12)}A^i \right)\right\}\\
	34\rho_2S^{\alpha}_{\alpha U}\left(\mathbf{e}_{(22)}U^i + \mathbf{e}_{(12)}A^i\right) + 34\rho_1S^{\alpha}_{\alpha A}\left(\mathbf{e}_{(22)}U^i + \mathbf{e}_{(12)}A^i\right) + \sqrt{2}\left(\mathbf{e}_{(23)} - i\mathbf{e}_{(24)}\right)[\beta_+S^{i0}_+ V_0^- +\\ +\beta_+S^{ij}_+V_j^- \left(\beta_+ + 17\rho_+\right)S^{\alpha}_{\alpha +}V^{i-}] + \sqrt{2}\left(\mathbf{e}_{(23)} + i\mathbf{e}_{(24)}\right)[\beta_-S^{i0}_- V_0^+ +\beta_-S^{ij}_-V_j^+ \left(\beta_- + 17\rho_-\right)S^{\alpha}_{\alpha -}V^{i+}]
\end{multline}

The last pair of longitudinal Gauss and Amp$\grave{e}$re laws are associated with the massive positive charged photon. They are

For $V^{L-}_{\mu}$:

Similarly, two longitudinal equations are obtained. They are
\begin{multline}
	\partial^{0}\left\{s_{-}S^{\alpha}_{\alpha -} + c_{-}\left(s^{\alpha}_{\alpha -A} + s^{\alpha}_{\alpha -U}\right)\right\}-\mathbf{m}^{2}_{V}V^{0-}_{L} + l^{0-}_{VL} + c^{0-}_{VL} = j^{0-}_{VL}
\end{multline} 
with
\begin{equation}
	l^{0-}_{VL} = -34\left(\mathbf{e}_{(33)} + \mathbf{e}_{(44)}\right)\left(s^{\alpha}_{\alpha AA} + s^{\alpha}_{\alpha UU}\right)V^{0-} -16\left(\mathbf{e}_{(33)}-\mathbf{e}_{(44)}\right)s^{\alpha}_{\alpha--}V^{0+}
	-32i\mathbf{e}_{(34)}s^{\alpha}_{\alpha--}V^{0+}
\end{equation}
\begin{multline}
	c^{0-}_{VL} = -\sqrt{2}\left(\mathbf{e}_{(13)}+i\mathbf{e}_{(14)}\right)\left[s^{00}_{-A}A_{0} + s^{0i}_{-A}A_{i}+17\left(s^{\alpha}_{\alpha -A} + s^{\alpha}_{\alpha -U}\right)A^{0} +s^{0i}_{-U}A_{i}\right]\\
	-\sqrt{2}\left(\mathbf{e}_{(23)}+i\mathbf{e}_{(24)}\right)\left[s^{00}_{-A}U_{0}+s^{0i}_{-A}U_{i} + 17\left(s^{\alpha}_{\alpha -A} + s^{\alpha}_{\alpha -U}\right)U^{0} + s{0i}_{-U}U_{i}\right],
\end{multline}
\begin{multline}
	j^{0-}_{VL} = \left(\mathbf{e}_{33} + \mathbf{e}_{(44)}\right)\left[\left( \beta_1\vec{S}_{A} + \beta_2 \vec{S}_U\right)\cdot \vec{V}^{-} + \left(\beta_1S^{\alpha}_{\alpha A}  + \beta_2S^{\alpha}_{\alpha U} + 17\rho_1S^{\alpha}_{\alpha A} + \rho_2S^{\alpha}_{\alpha U} \right)\phi^{-}\right]\\
	+\sqrt{2}\left(\mathbf{e}_{(13)} + i \mathbf{e}_{(14)}\right)\left[\beta_- \vec{S}_{-}\cdot \vec{A} +  \left(\beta_- + 17 \rho_- \right)S^{\alpha}_{\alpha -}\phi_A\right] +\sqrt{2}\left(\mathbf{e}_{(23)} + i \mathbf{e}_{(24)}\right)[\beta_- \vec{S}_{-}\cdot \vec{U} +  \\
	+\left(\beta_- + 17 \rho_- \right)S^{\alpha}_{\alpha -}\phi_U]
\end{multline}
and
\begin{equation}
	\partial^{i}\left\{s_{-}S^{\alpha}_{\alpha -} + c_{-}\left(s^{\alpha}_{\alpha -A} + s^{\alpha}_{\alpha -U}\right)\right\} -\mathbf{m}^{2}_{V}V^{i-}_{L} + l^{i-}_{VL} + c^{i-}_{VL} = j^{i-}_{VL}
\end{equation}
with
\begin{multline}
	l^{i-}_{VL} = -\left(\mathbf{e}_{(33)} + \mathbf{e}_{(44)}\right)\left(s^{\alpha}_{\alpha AA} + s^{\alpha}_{\alpha UU}\right)V^{i-} -16\left(\mathbf{e}_{(33)}-\mathbf{e}_{(44)}\right)s^{\alpha}_{\alpha --}V^{i+} - 32\mathbf{e}_{(34)}s^{\alpha}_{\alpha --}V^{i+}
\end{multline}
\begin{multline}
	c^{i-}_{VL} = -\sqrt{2}\left(\mathbf{e}_{(13)} +i \mathbf{e}_{(14)}\right)\left[\left(s^{0i}_{-A} + s^{0i}_{-U}\right)A_0 + \left(s^{ij}_{-A} + s^{ij}_{-U}\right)A_{j} + 17\left(s^{\alpha}_{\alpha -A} + s^{\alpha}_{\alpha -U}\right)A^{i}\right]+\\
	-\sqrt{2}\left(\mathbf{e}_{(23)} +i \mathbf{e}_{(24)}\right)\left[\left(s^{0i}_{-A} + s^{0i}_{-U}\right)U_{0}  + \left(s^{ij}_{-A} +s^{ij}_{-U}\right)U_{j} + 17\left(s^{\alpha}_{\alpha -A} + s^{\alpha}_{\alpha -U}\right)U^{i} \right]+\\
	-2\left(\mathbf{e}_{(33)} + \mathbf{e}_{(44)} \right)\left[\left(s^{0i}_{AA}+s^{0i}_{AU} + s^{0i}_{UU} + s^{0i}_{+-}\right)V^{-}_0 + \left(s^{ij}_{AA} + s^{ij}_{AU} + s^{ij}_{UU} + s^{ij}_{+-}\right)V^{-}_{j} + \left(s^{\alpha}_{\alpha AU} + s^{\alpha}_{\alpha +-}\right)V^{i-}\right]\\
	\left(\mathbf{e}_{(33)} - \mathbf{e}_{(44)}\right)s^{\alpha}_{\alpha --}V^{i-}-2i\mathbf{e}_{(34)}s^{ij}_{--}V^{+}_{j}
\end{multline}

\begin{multline}
	j^{i-}_{VL} = \left(\mathbf{e}_{33} + \mathbf{e}_{(44)}\right)\left[\left(\beta_1S_A^{0i} + \beta_2S_U^{0i}\right)V_0{+} + \left(\beta_1S^{\alpha}_{\alpha}\right)V^{i+} + \left(\rho_1S^{\alpha}_{\alpha A} + \rho_2S^{\alpha}_{\alpha U}\right)V^{i+}\right]+\\
	+\sqrt{2}\left(\mathbf{e}_{(13)} - i \mathbf{e}_{(14)}\right)\left[\beta_+S^{0i}_+A_0 + \beta_+S^{ji}_+A_j + \left(\beta_+ + 17 \rho_+\right)S^{\alpha}_{\alpha +}A^{i}\right] +\\
	+\sqrt{2}\left(\mathbf{e}_{(23)} - i \mathbf{e}_{(24)}\right)\left[\beta_+S^{0i}_+U_0 + \beta_+S^{ji}_+U_j + \left(\beta_+ + 17 \rho_+\right)S^{\alpha}_{\alpha +}U^{i}\right] +\\
	-\beta_-\left(\mathbf{e}_{(13)} + i\mathbf{e}_{(14)}\right)\left\{\frac{1}{\sqrt{2}}\left[S^{\alpha}_{\alpha +}A^{i} + S^{\alpha}_{\alpha A}V^{i+}\right]+\sqrt{2}\left[S_+^{0i}A_0 + S_+^{ji}A_j + S^{0i}_AV_0^{+} + S^{ji}_A V^{i+}\right]\right\}\\
	-\beta_+\left(\mathbf{e}_{(23)} + i\mathbf{e}_{(24)}\right)\left\{\frac{1}{\sqrt{2}}\left[S^{\alpha}_{\alpha +}U^{i} + S^{\alpha}_{\alpha U}V^{i+}\right]+\sqrt{2}\left[S_+^{0i}U_0 + S_+^{ji}U_j + S^{0i}_UV_0^{+} + S^{ji}_U V^{i+}\right]\right\}
\end{multline}
For $V^{L+}_{\mu}$:
\begin{equation}
	\partial^{0}\left\{s_{+}S^{\alpha}_{\alpha +} + c_{+}\left(s^{\alpha}_{\alpha +A} + s^{\alpha}_{\alpha +U}\right)\right\}-\mathbf{m}^{2}_{V}V^{0+}_{L} + l^{0+}_{VL} + c^{0+}_{VL} = j^{0+}_{VL}
\end{equation}
with
\begin{multline}
	l^{0+}_{VL} = -34\left(\mathbf{e}_{(13)}-i\mathbf{e}_{(14)}\right)\left(s^{\alpha}_{\alpha +A} + s^{\alpha}_{\alpha UU} \right)V^{0+} -16\left(\mathbf{e}_{(33)}-\mathbf{e}_{(44)}\right)s^{\alpha}_{\alpha ++}V^{0-}-32\mathbf{e}_{(34)}s^{\alpha}_{\alpha ++}V^{0-},
\end{multline}
\begin{multline}
	c^{0+}_{VL} = -\sqrt{2}\left(\mathbf{e}_{(13)}-i\mathbf{e}_{(14)}\right)\left[\left(s^{00}_{+A} + s^{00}_{+U} \right)A_{0} + \left(s^{0i}_{+A} + s^{0i}_{+U}\right)A_{i} + 17\left(s^{\alpha}_{\alpha +A} + s^{\alpha}_{\alpha +U}\right)A^{0}\right]\\
	-\sqrt{2}\left(\mathbf{e}_{(24)} - i\mathbf{e}_{(24)}\right)\left[\left(s^{00}_{+A} + s^{00}_{+U}\right)U_{0} + \left(s^{0i}_{+A} + s^{0i}_{+U}\right)U_{i} + \left(s^{\alpha}_{\alpha +A} + s^{\alpha}_{\alpha +U}\right)U^{0}\right]\\-2\left(\mathbf{e}_{(33)} + \mathbf{e}_{(44)}\right)\left[\left(s^{00}_{AA} + s^{00}_{UU} + s^{00}_{AU} + s^{00}_{+-}\right)V_{0}^{+} + \left(s^{0i}_{AA} + s^{0i}_{UU} + s^{0i}_{AU} + s^{0i}_{+-}\right)V^{+}_{i} + s^{\alpha}_{\alpha +-}V^{0+}\right]\\
	-\left(\mathbf{e}_{(33)} - \mathbf{e}_{(44)}\right)s^{\alpha}_{\alpha ++}V^{0-} - 2i\mathbf{e}_{(34)}s^{00}_{++}V^{-}_{0}-2i\mathbf{e}_{(34)}s^{0i}_{++}V^{-}_{i},
\end{multline}
\begin{multline}
	j^{0+}_{VL} = \left(\mathbf{e}_{33} + \mathbf{e}_{(44)}\right)\left[\left( \beta_1S^{0i}_{A} + \beta_2 S^{0i}_U\right)\cdot V_{i}^{+} + \left(\beta_1S^{\alpha}_{\alpha A}  + \beta_2S^{\alpha}_{\alpha U} + 17\rho_1S^{\alpha}_{\alpha A} + \rho_2S^{\alpha}_{\alpha U} \right)V^{0+}\right]\\
	+\sqrt{2}\left(\mathbf{e}_{(13)} - i \mathbf{e}_{(14)}\right)\left[\beta_+ S^{0i}_{+}A_{i} +  \left(\beta_+ + 17 \rho_+ \right)S^{\alpha}_{\alpha +}A_{0}\right] +\sqrt{2}\left(\mathbf{e}_{(23)} - i \mathbf{e}_{(24)}\right)[\beta_+ S^{0i}_{+}U_{i} +  \\
	+\left(\beta_+ + 17 \rho_+ \right)S^{\alpha}_{\alpha +}U_{0}].
\end{multline}
and
\begin{equation}
	\partial^{i}\left[s_{+}S^{\alpha}_{\alpha +} + c_{+}\left(s^{\alpha}_{\alpha +A} + s^{\alpha}_{\alpha +U}\right)\right] -\mathbf{m}^{2}_{V}V^{i+}_{L} + l^{i+}_{VL} + c^{i+}_{VL} = j^{i+}_{VL}
\end{equation}
with
\begin{multline}
	l^{i+}_{VL} = -34\left(\mathbf{e}_{(33)} + \mathbf{e}_{(44)}\right)\left(s^{\alpha}_{\alpha AA} + s^{\alpha}_{\alpha UU}\right)V^{i+} - 16\left(\mathbf{e}_{(33)} - \mathbf{e}_{(44)}\right)s^{\alpha}_{\alpha ++}V^{i-}-32i\mathbf{e}_{(34)}s^{\alpha}_{\alpha ++}V^{i-},
\end{multline}
\begin{multline}
	c^{i+}_{VL} = -\sqrt{2}\left(\mathbf{e}_{(13)} - i\mathbf{e}_{(14)}\right)\left[\left(s^{0i}_{+A} + s^{0i}_{+U}\right)A_{0} + \left(s^{ij}_{+A} + s^{ij}_{+U}\right)A_{j} + 17\left(s^{\alpha}_{\alpha +A} + s^{\alpha}_{\alpha +U} \right)A^{i}\right]\\
	-\sqrt{2}\left(\mathbf{e}_{(23)} - i\mathbf{e}_{(24)}\right)\left[\left(s^{0i}_{+A} + s^{0i}_{+U}\right)U_{0} + \left(s^{ij}_{+A} + s^{ij}_{+U}\right)U_{j} + 17\left(s^{\alpha}_{\alpha +A} + s^{\alpha}_{\alpha +U} \right)U^{i}\right]\\
	-2\left(\mathbf{e}_{(33)} +\mathbf{e}_{(44)}\right)\left[\left(s^{0i}_{AA} + s^{0i}_{UU} + s^{0i}_{AU} + s^{0i}_{+-}\right)V^{+}_{0} + \left(s^{ij}_{AA} + s^{ij}_{UU} + s^{ij}_{AU} + s^{ij}_{+-}\right)V^{+}_{j} + \left(s^{\alpha}_{\alpha AU} + s^{\alpha}_{\alpha +-}\right)V^{i+}\right]\\
	-\left(\mathbf{e}_{(33)} - \mathbf{e}_{(44)}\right)s^{\alpha}_{\alpha ++}V^{i-} -2i\mathbf{e}_{(34)}s^{0i}V^{-}_{0}-2i\mathbf{e}_{(34)}s^{ij}_{++}V_{j}^{-},
\end{multline}
\begin{multline}
	\vec{j}^{i+}_{V L}=\left(\mathbf{e}_{33} + \mathbf{e}_{(44)}\right)\left[\left(\beta_1S_A^{0i} + \beta_2S_U^{0i}\right)V_0{+} + \left(\beta_1S^{\alpha}_{\alpha}\right)V^{i+} + \left(\rho_1S^{\alpha}_{\alpha A} + \rho_2S^{\alpha}_{\alpha U}\right)V^{i+}\right]+\\
	+\sqrt{2}\left(\mathbf{e}_{(13)} - i \mathbf{e}_{(14)}\right)\left[\beta_+S^{0i}_+A_0 + \beta_+S^{ji}_+A_j + \left(\beta_+ + 17 \rho_+\right)S^{\alpha}_{\alpha +}A^{i}\right] +\\
	+\sqrt{2}\left(\mathbf{e}_{(23)} - i \mathbf{e}_{(24)}\right)\left[\beta_+S^{0i}_+U_0 + \beta_+S^{ji}_+U_j + \left(\beta_+ + 17 \rho_+\right)S^{\alpha}_{\alpha +}U^{i}\right]\\
	-\beta_-\left(\mathbf{e}_{(13)} + i\mathbf{e}_{(14)}\right)\left\{\frac{1}{\sqrt{2}}\left[S^{\alpha}_{\alpha +}A^{i} + S^{\alpha}_{\alpha A}V^{i+}\right]+\sqrt{2}\left[S_+^{0i}A_0 + S_+^{ji}A_j + S^{0i}_AV_0^{+} + S^{ji}_ A V^{i+}\right]\right\}\\
	-\beta_+\left(\mathbf{e}_{(23)} + i\mathbf{e}_{(24)}\right)\left\{\frac{1}{\sqrt{2}}\left[S^{\alpha}_{\alpha +}U^{i} + S^{\alpha}_{\alpha U}V^{i+}\right]+\sqrt{2}\left[S_+^{0i}U_0 + S_+^{ji}U_j + S^{0i}_UV_0^{+} + S^{ji}_U V^{i+}\right]\right\}.
\end{multline}
Eqs. (4.53-4.69) will follow the same conservation law as eq. (4.46).

A four-four electromagnetism is generated at fundamental level. The above equations respect originals EM postulates and expand the EM behaviour. Introducing as EM completeness, four interconnected photons. A new EM dynamics is derived. It contains not only nonlinear granular, as collective EM fields and the presence of potential fields. Adimensional coupling constants $g_{I}$ between EM fields and vector potential fields  are introduced. not depending on electric charge. Mass without the Higgs mechanism is incorporated. Two physics appear with spin-1 and spin-0, differing, as observables and dynamics, 

Thus, the above equations are showing how symmetry is more important than nature constants. The electric charge universality appears through its symmetry and no more as a coupling constant. These equations are introducing diverse coupling constants without depending on electric charge. Expressing that, the EM principle is on electric charge symmetry, and not, on its Millikan value.

\section{Collective Bianchi identities}

EM quadruplet introduces collective Bianchi identities with sources. They are associated with each collective vector bosons fields. It gives antisymmetric and symmetric Bianchi Identities.

For antisymmetric sector:
\begin{equation}
	\vec{\nabla} \times \vec{\mathbf{e}}_{AU} + \frac{\partial}{\partial t}\vec{b}_{AU} = \mathbf{e}_{[12]}\left(\vec{A} \times \vec{E}_U - \vec{U}\times \vec{E}_A + \phi_A\vec{B}_U - \phi_U\vec{B}_A\right)
\end{equation}
\begin{equation}
	\vec{\nabla}\cdot \vec{b}_{AU} = \mathbf{e}_{[12]}\left(-\vec{A}\cdot \vec{B}_U - \vec{U} \cdot \vec{B}_A\right),
\end{equation}
with the following conservation law.
\begin{equation}
	\frac{\partial}{\partial t}\left(\vec{A}\cdot\vec{B}_{U}+\vec{U}\cdot
	\vec{B}_{A}\right) = \vec{\nabla}\cdot\left(\vec{A}\times\vec{E}_{U} - \vec{U}\times\vec{E}_{A} + \phi_{A}\vec{B}_{U} - \phi_{U}\vec{B}_{A}\right)
\end{equation}

Thus, the magnetic monopole comes out naturally from the extended abelian symmetry. A structure similar to ice glasses is expressed in terms of fields [35].

\begin{equation}
	\vec{\nabla} \times \vec{\mathbf{e}_{+-}} + \frac{\partial}{\partial t} \vec{b}_{+-} = -i \mathbf{e}_{[34]}Im\left\{\vec{V}^{+} \times \vec{E}_{-} - \phi^{+}\vec{B}_-\right\}
\end{equation}
\begin{equation}
	\vec{\nabla}\cdot \vec{b}_{+-} = i\mathbf{e}_{[+4]}Im\left\{\vec{V}^{+}\cdot \vec{B}_-\right\},
\end{equation}
\begin{eqnarray}
	&&\vec{\nabla} \times \left(\vec{\mathbf{e}}_{+A} +\vec{\mathbf{e}}_{-A}\right) + \frac{\partial}{\partial t}\left(\vec{b}_{+A} + \vec{b}_{-A}\right) = Re\big\{\frac{1}{\sqrt{2}}\left(\mathbf{e}_{[13]} + i\mathbf{e}_{[14]}\right)\big(\vec{A}\ \times \vec{E}_+\nonumber
	\\
	&&- \vec{V}^+ \times \vec{E}_{A} + \phi_A\vec{B}_+ - \phi_+ \vec{B}_A\big)\big\}
\end{eqnarray}
\begin{equation}
	\vec{\nabla} \cdot \left(\vec{b}_{+A} + \vec{b}_{-A}\right) = Re\left\{\frac{1}{\sqrt{2}} \left(\mathbf{e}_{[13]} + i \mathbf{e}_{[14]}\right)\left(-\vec{A} \cdot \vec{B}_+ +\vec{V}^{+}\cdot \vec{B}_A\right)\right\},
\end{equation}
\begin{eqnarray}
	&&\vec{\nabla} \times \left(\vec{\mathbf{e}}_{+U} +\vec{\mathbf{e}}_{-U}\right) + \frac{\partial}{\partial t}\left(\vec{b}_{+U} + \vec{b}_{-U}\right) = Re\big\{\frac{1}{\sqrt{2}}\left(\mathbf{e}_{[23]} + i\mathbf{e}_{[24]}\right)\big(\vec{U}\ \times \vec{E}_++\nonumber
	\\
	 &&- \vec{V}^+ \times \vec{E}_{U} + \phi_U\vec{B}_+ - \phi_+ \vec{B}_U\big)\big\}
\end{eqnarray}
\begin{equation}
	\vec{\nabla} \cdot \left(\vec{b}_{+U} + \vec{b}_{-U}\right) = Re\left\{\frac{1}{\sqrt{2}} \left(\mathbf{e}_{[23]} + i \mathbf{e}_{[24]}\right)\left(-\vec{U} \cdot \vec{B}_+ +\vec{V}^{+}\cdot \vec{B}_U\right)\right\},
\end{equation}

Eq. (5.4-5.9) will follow the same conservation law as eq. (5.3)

For symmetric sector:
\begin{equation}
	\partial^{i}s^{0j}_{AA} + \partial^{j}s^{i0}_{AA} + \partial^{0}s^{ij}_{AA} = \mathbf{e}_{(11)}\left\{A^{i}S^{0j}_{A} + A^{j}S^{0i}_{A} + A^{0}S^{ij}\right\}
\end{equation}
\begin{equation}
	\frac{\partial}{\partial t}\vec{s}_{AA} = \mathbf{e}_{(11)}\left\{\phi_{A}\cdot \vec{S}_{A} + \vec{A}S_{A}\right\},
\end{equation}
where eqs. (5.9) and (5.10) are showing the photon Faraday law, Similarly for other quadruplet fields
\begin{equation}
	\partial^{i}s^{0j}_{UU} + \partial^{j}s^{i0}_{UU} + \partial^{0}s^{ij}_{UU} = \mathbf{e}_{(22)}\left\{U^{i}S^{0j}_{U} + U^{j}S^{0i}_{U} + U^{0}S^{ij}\right\}
\end{equation}
\begin{equation}
	\frac{\partial}{\partial t}\vec{s}_{UU} = \mathbf{e}_{(11)}\left\{\phi_{U}\cdot \vec{S}_{U} + \vec{U}S_{U}\right\},
\end{equation}

\begin{equation}
	\partial^{i}s^{0j}_{AU} + \partial^{j}s^{i0}_{AU} + \partial^{0}s^{ij}_{AU} = \mathbf{e}_{(12)}\left\{A^{i}S^{0j}_{U} + A^{j}S^{0i}_{U} + A^{0}S^{ij}_{U} + U^{i}S^{0j}_{A} + U^{j}S^{0i}_{A} + U^{0}S^{ij}_{A} \right\}
\end{equation}
\begin{equation}
	\frac{\partial}{\partial t}\vec{s}_{AU} = \mathbf{e}_{(12)}\left\{\phi_{A} \vec{S}_{U} + \vec{A}S_{U} + \phi_{U} \vec{S}_{A} + \vec{U}S_{A}\right\},
\end{equation}

\begin{equation}
	\partial^{i}s^{0j}_{+-} + \partial^{j}s^{i0}_{+-} + \partial^{0}s^{ij}_{+-} = \left(\mathbf{e}_{(33)} + \mathbf{e}_{(44)}\right)Re\left\{V^{i+}S^{0j}_{-} + V^{j+}S^{0i}_{-} + A^{0}S^{ij}_-\right\}
\end{equation}
\begin{equation}
	\frac{\partial}{\partial t}\vec{s}_{+-} = \left(\mathbf{e}_{(33)} + \mathbf{e}_{(44)}\right)Re\left\{\phi_{+}\vec{S}_{+} + \vec{V}^{+}S_{-}\right\},
\end{equation}

\begin{eqnarray}
	&&\partial^{i}\left(s^{0j}_{A+} + s^{0j}_{A-} \right)+ \partial^{j}\left(s^{i0}_{A+} + s^{i0}_{A-} \right) \partial^{0}\left(s^{ij}_{A+} + s^{ij}_{A-}+ \right) = 
	\\
	&&\left(\mathbf{e}_{(13)} + i\mathbf{e}_{(14)}\right)Re\{V^{i+}S^{0j}_{A} +\nonumber V^{j+}S^{0i}_{A} +V^{0+}S^{ij}_A + A^{i}S^{0j}_{+} + A^{j}S^{0i}_{+} + A^{0}S^{ij}_+\}
\end{eqnarray}
\begin{eqnarray}
	\frac{\partial}{\partial t}\left(\vec{s}_{A+} + \vec{s}_{A-} \right) = \left(\mathbf{e}_{(13)} + i\mathbf{e}_{(14)}\right)Re\left\{\phi_{+}\vec{S}_{A} + \vec{V}^{+}S_{A} + \phi_{A}\vec{S}_{+} + \vec{A}S_{+}\right\},
\end{eqnarray}

\begin{eqnarray}
	&&\partial^{i}\left(s^{0j}_{U+} + s^{0j}_{U-} \right)+ \partial^{j}\left(s^{i0}_{U+} + s^{i0}_{U-} \right) \partial^{0}\left(s^{ij}_{U+} + s^{ij}_{U-}+ \right) 
	\\
	&&= \left(\mathbf{e}_{(13)} + i\mathbf{e}_{(14)}\right)Re\{V^{i+}S^{0j}_{U} + V^{j+}S^{0i}_{U} +
	V^{0+}S^{ij}_U + U^{i}S^{0j}_{+} + U^{j}S^{0i}_{+} + U^{0}S^{ij}_+\}\nonumber
\end{eqnarray}
\begin{equation}
	\frac{\partial}{\partial t}\left(\vec{s}_{U+} + \vec{s}_{U-} \right) = \left(\mathbf{e}_{(13)} + i\mathbf{e}_{(14)}\right)Re\left\{\phi_{+}\vec{S}_{U} + \vec{V}^{+}S_{U} + \phi_{U}\vec{S}_{+} + \vec{U}S_{+}\right\},
\end{equation}

\begin{eqnarray}
	&&\partial^{i}\left(s^{0j}_{++} + s^{0j}_{--} \right)+ \partial^{j}\left(s^{i0}_{++} + s^{i0}_{--} \right) \partial^{0}\left(s^{ij}_{++} + s^{ij}_{--}+ \right) = 
	\\
	&&\left(\mathbf{e}_{(33)} - \mathbf{e}_{(44)}\right)\frac{1}{2}\{V^{i+}S^{0j}_{-} + V^{j+}S^{0i}_{-} +V^{0+}S^{ij}_- + V^{i-}S^{0j}_{+} + V^{j-}S^{0i}_{+} + V^{0-}S^{ij}_+\}\nonumber
\end{eqnarray}
\begin{equation}
	\frac{\partial}{\partial t}\left(\vec{s}_{++} + \vec{s}_{--} \right) = \left(\mathbf{e}_{(33)} - \mathbf{e}_{(44)}\right)\left\{\phi_{+}\vec{S}_{+} + \vec{V}^{+}S_{+} + \phi_{+}\vec{S}_{+} + \vec{V}^+S_{+}\right\},
\end{equation}

\begin{eqnarray}
	&&\partial^{i}\left(s^{0j}_{++34} + s^{0j}_{--34} \right)+ \partial^{j}\left(s^{i0}_{++34} + s^{i0}_{--34} \right) \partial^{0}\left(s^{ij}_{++34} + s^{ij}_{--34}+ \right) =
	\\
   && i\mathbf{e}_{(34)}Im\{V^{i+}S^{0j}_{+} + V^{j+}S^{0i}_{+} + V^{0+}S^{ij}_+\}\nonumber 
\end{eqnarray}
\begin{equation}
	\frac{\partial}{\partial t}\left(\vec{s}_{++} + \vec{s}_{--} \right) = i\mathbf{e}_{(34)}Im\left\{\phi_{+}\vec{S}_{+} + \vec{V}^{+}S_{+} \right\}.
\end{equation}

The above equations are introducing a new physicality in theory. Expanding the presence of electromagnetic induction. New Faraday laws with monopoles are obtained. Notice that the symmetric sector does not provide conservation laws.

\section{Final Considerations}

A fundamental EM beyond Maxwell is required. Nonlinearity [36], strong magnetic fields [37], and photonics [38] are introducing new phenomenologies for electromagnetism to be extended. There is a more fundamental EM to be excavated under electric charge symmetry. Consider Maxwell just as an EM sector. Discover new electric and magnetic fields and corresponding equations.
 
A constitutive four photons dynamics is proposed. A new significance for electric charge symmetry appears. A quadruplet EM completeness emerges from charge transfer. The four bosons electromagnetism deploys new aspects of electric charge, light, and spin. Electric charge physics is more than stipulate Maxwell EM fields, continuity equation, and coupling constant. A fundamental EM theory must be supported by a fundamental electric charge symmetry. It associates the four intermediate gauge bosons. It arranges the field's quadruplet $\{A_{\mu}, U_{\mu}, V_{\mu}^{\pm}\}$ as homothety for triangle sides. New relationships between electric charge symmetry and EM fields are produced. Through the correspondent gauge homothety, it constitutes the Lagrangian [22-23], generating new EM observables described in section 1 and three Noether identities in section 3.

The second argument for four bosons electromagnetism to be a candidate for a fundamental EM is on primordial light. A new nature for light is proposed. A constitutive light associated with three other intermediate bosons is performed. A primitive photon is encountered. The four bosons EM provides a light more primordial than being just a Maxwell wave. The Lagrangian study introduces physics that includes sectors from Maxwell to photonics. It contains light invariance, ubiquous and selfinteracting photons. Photonics is derived. Photon acting as own source, generating granular and collective fields strengths, interconnected, sharing a quadruplet photon physics and selfinteracting at tree level. An inductive photon Faraday law is proposed, as eqs (2.6,5.10-5.11) shows. A Lorentz force depending on the photon field is expressed [22]. Feynman's vertices with tri-and-quadrilinear are obtained [39]. Pure photonics is constituted, The third aspect is spin. From the heuristic Stern-Gerlach experiment [40], spin is incorporated on fields formalism as [24].
 
A methodology to analyse a model's significance is how far it is incorporated into the historical process. As relativity came out as an extension of Newtonian mechanics, a new EM model should happen inserted into the EM development. The EM process may be viewed in three historical phases. First, through charges and magnets, as described by William Gilbert in 1600 [41]; second, charges and fields, by Maxwell in 1864; and third, on fields and fields, being expected at this 21$^{th}$ century.

As Faraday, the four bosons, EM penetrates in the fields-fields region. A pure EM field's physics is expressed. Surpass Maxwell's limitations, the myth of matter and light as EM waves. The first aspect was treated in the introduction. The myth of matter appears when we look at solid objects around us and ask what is going on. The perspective is to envisage matter concepts as guiding nature. Then, it is noticed the presence of an empty space filled with fields. Something says that nature is not only made by moving particles but also through field dynamics. Faraday was the first one to perceive this physics beyond matter. An enlargement of Faraday's perception is expected from an EM development.  

The four bosons EM nonlinearity advances the concept of matter depending on fields. Following Faraday, introduces a view where physical laws are beyond matter. As a premise from electric charge symmetry. Consider fields as the most fundamental objects to construct the world. The concept of matter derived from fields. Rewrite the relationships between matter and fields by developing fields grouping, nonlinear fields, fields being own sources, mass and electric charge depending on fields, diverse Faraday laws, fields monopoles, Lorentz force depending on potential fields. Expressing also electric charge and mass in terms of continuity equations depending on fields, as in section 3.
 
A consequence of matter depending on fields is the possibility of reinterpreting on dark matter and dark energy in terms of field properties. Considering that the corresponding energy-momentum produces a negative pressure depending just on scalar potential fields [22], it may be a candidate for dark energy. The diagond term $T_{ii}$ be responsible for universe expansion [42].

Fundamental electromagnetism is proposed. A three charges EM is performed. An EM completeness by four intermediate photons is established beyond QED [43]. A new EM energy is discovered. New EM regimes are obtained. Overall, seven interrelated EM sectors are developed based on the presence of new EM observables. They are Maxwell, systemic, nonlinear, neutral, spintronics, electroweak, and photonics.
%

\begin{appendices}
	
	\section{Apendice: densitities and currents at section 2}
	
	For $U_{\mu}$:
	
	Spin-1:
	\begin{multline}
		\rho^T_{U} \equiv -2\mathbf{e}_{[12]}\left(\vec{E}_A + \vec{E}_U\right) \cdot \vec{A} - \sqrt{2}\left(\mathbf{e}_{[23]} - i\mathbf{e}_{[24]}\right)\vec{E}_+\cdot \vec{V}^- - \sqrt{2}\left(\mathbf{e}_{[23]} + i\mathbf{e}_{[24]}\right)\vec{E}_-\cdot \vec{V}^+ +\\
		-4\mathbf{e}_{[12]} \left(\vec{\mathbf{e}}_{AU} + \vec{\mathbf{e}}_{+-} \right) \cdot \vec{A}  -\sqrt{2}\left(\mathbf{e}_{[23]} - i\mathbf{e}_{[24]}\right) \left(\vec{\mathbf{e}}_{A+} + \vec{\mathbf{e}}_{U+} \right) \cdot \vec{V}^- - \sqrt{2}\left(\mathbf{e}_{[23]} + i\mathbf{e}_{[24]}\right) \left(\vec{\mathbf{e}}_{A-} + \vec{\mathbf{e}}_{U-} \right) \cdot \vec{V}^+
	\end{multline}
and

\begin{multline}
	\vec{J}^T_U \equiv -2 \mathbf{e}_{[12]}\left[\left(\vec{E}_A + \vec{E}_U\right)\phi_A + \left(\vec{B}_A + \vec{B}_U\right) \times \vec{A}\right] - \sqrt{2}\left(\mathbf{e}_{[23]}-i\mathbf{e}_{[24]}\right)\left[\vec{E}_+ \phi^- + \vec{B}_+ \times \vec{V}^-\right] -\\ \sqrt{2}\left(\mathbf{e}_{[23]}+i\mathbf{e}_{[24]}\right)\left[\vec{E}_- \phi^+ + \vec{B}_- \times \vec{V}^+\right] - 4 \mathbf{e}_{[12]}\left[\vec{\mathbf{e}}_{AU}\phi_A + \left(\vec{b}_{AU} \times \vec{A}\right)\right] \\ -\sqrt{2}\left(\mathbf{e}_{[23]}-i\mathbf{e}_{[24]}\right)\left[ \left(\vec{\mathbf{e}}_{+A} +\vec{\mathbf{e}}_{+U} \right) \phi^- + \left(\vec{b}_{+A} + \vec{b}_{+U} \right)\times \vec{V}^- \right] +\\
	-\sqrt{2}\left(\mathbf{e}_{[23]}+i\mathbf{e}_{[24]}\right)\left[ \left(\vec{\mathbf{e}}_{-A} +\vec{\mathbf{e}}_{-U} \right) \phi^+ + \left(\vec{b}_{-A} + \vec{b}_{-U} \right)\times \vec{V}^+\right]
\end{multline}

For spin-0:
\begin{multline}
	\rho^{s}_{U} \equiv 2 \left(\mathbf{e}_{(22)} + \mathbf{e}_{(12)}\right) \left\{ \left(\beta_1\vec{S}_A + \beta_2 \vec{S}_U \right) \cdot \left(\vec{A} + \vec{U}\right)  + \left[\left( \beta_1 + 17\rho_1 \right)S^{\alpha}_{\alpha A} + \left(\beta_2 +17\rho_2\right) S^{\alpha}_{\alpha U} \right] \left(\phi_A + \phi_U\right) \right\}\\
	+\sqrt{2}\left(\mathbf{e}_{(23)} - i\mathbf{e}_{(24)}\right)\left[\beta_+ \vec{S}^+\vec{V}^- + \left(\beta_+ + 17\rho_+\right)S^{\alpha}_{\alpha +}\phi^-\right] - \beta_1\mathbf{e}_{(22)}\left(S^{\alpha}_{\alpha U} \phi_U + 2 \vec{S}_U \cdot \vec{U}\right)+\\
	-\beta_2\mathbf{e}_{(12)}\left(S^{\alpha}_{\alpha A}\phi_U + S^{\alpha}_{\alpha U}\phi_A + \vec{S}_A \cdot \vec{U} + \vec{S}_U \cdot \vec{A}\right) - \beta_2\left(\mathbf{e}_{(33)} + \mathbf{e}_{(44)}\right) \left(S^{\alpha}_{\alpha +}\phi^- + S^{\alpha}_{\alpha -}\phi^+ + \vec{S}_+\vec{V}^- + \vec{S}_-\vec{V}^+\right)\\
	+\sqrt{2}\left(\mathbf{e}_{(23)} + i\mathbf{e}_{(24)}\right)\left[\beta_- \vec{S}^-\vec{V}^+ + \left(\beta_- + 17\rho_-\right)S^{\alpha}_{\alpha -}\phi^+\right] + 4\mathbf{e}_{(22)}\left(\vec{s}_{AA} + \vec{s}_{UU} + \vec{s}_{AU} +\vec{s}_{+-} \right) \cdot \vec{U}\\
	+72\mathbf{e}_{(22)}\left(s^{\alpha}_{\alpha AA} + s^{\alpha}_{\alpha AU} + s^{\alpha}_{\alpha+-}\right)\phi_U + 4 \mathbf{e}_{(12)} \left( \vec{s}_{AA} + \vec{s}_{UU} \right) \cdot \vec{A} + 72 \mathbf{e}_{(12)}\left(s^{\alpha}_{\alpha AA} + \mathbf{e}^{\alpha}_{\alpha UU} + s^{\alpha}_{+-}\right)\phi_A\\
	-\beta_1\mathbf{e}_{(22)}\left(\vec{S}_U\cdot \vec{U} + S^{\alpha}_{\alpha U} \right) + 
	+\sqrt{2}\left(\mathbf{e}_{(13)} + i\mathbf{e}_{(14)}\right) \left[\vec{s}_{-A}\vec{V}^+ + 18 s^{\alpha}_{\alpha A-} \phi^+\right] + \\
	\sqrt{2}\left(\mathbf{e}_{(13)} - i\mathbf{e}_{(14)}\right) \left[\vec{s}_{+A}\vec{V}^- + 18 s^{\alpha}_{\alpha A+} \phi^-\right]
	+\sqrt{2}\left(\mathbf{e}_{(23)} + i\mathbf{e}_{(24)}\right) \left[\vec{s}_{-U}\vec{V}^+ + 18 s^{\alpha}_{\alpha U-} \phi^+\right] +\\
	+ \sqrt{2}\left(\mathbf{e}_{(23)} - i\mathbf{e}_{(24)}\right) \left[\vec{s}_{+U}\vec{V}^- + 18 s^{\alpha}_{\alpha U+} \phi^-\right]
\end{multline}
and
\begin{multline}
	\vec{j}^{s}_U \equiv 2\beta_2\left\{S^{i0}_U\left(\mathbf{e}_{(22)}U_0 + \mathbf{e}_{(12)}A_0\right) + S^{ij}_U\left( \mathbf{e}_{(22)}U_j + \mathbf{e}_{(12)}A_j \right) + S^{\alpha}_{\alpha U}\left( \mathbf{e}_{(22)}U^i + \mathbf{e}_{(12)}A^i \right)\right\}\\
	+2\beta_1\left\{S^{i0}_A\left(\mathbf{e}_{(22)}U_0 + \mathbf{e}_{(12)}A_0\right) + S^{ij}_A\left( \mathbf{e}_{(22)}U_j + \mathbf{e}_{(12)}A_j \right) + S^{\alpha}_{\alpha A}\left( \mathbf{e}_{(22)}U^i + \mathbf{e}_{(12)}A^i \right)\right\}\\
	34\rho_2S^{\alpha}_{\alpha U}\left(\mathbf{e}_{(22)}U^i + \mathbf{e}_{(12)}A^i\right) + 34\rho_1S^{\alpha}_{\alpha A}\left(\mathbf{e}_{(22)}U^i + \mathbf{e}_{(12)}A^i\right) + \sqrt{2}\left(\mathbf{e}_{(23)} - i\mathbf{e}_{(24)}\right)[\beta_+S^{i0}_+ V_0^- +\\ +\beta_+S^{ij}_+V_j^- \left(\beta_+ + 17\rho_+\right)S^{\alpha}_{\alpha +}V^{i-}] + \sqrt{2}\left(\mathbf{e}_{(23)} + i\mathbf{e}_{(24)}\right)[\beta_-S^{i0}_- V_0^+ +\beta_-S^{ij}_-V_j^+ \left(\beta_- + 17\rho_-\right)S^{\alpha}_{\alpha -}V^{i+}]\\
	+4\mathbf{e}_{(22)}[\left(s^{i0}_{AA} + s^{i0}_{UU} + s^{i0}_{AU} + s^{i0}_{+-} \right)A_0 + \left(s^{ij}_{AA} + s^{ij}_{UU} + s^{ij}_{AU} + s^{ij}_{+-} \right)U_j + (2s^{\alpha}_{\alpha AA} + 17s^{\alpha}_{\alpha UU} + 17s^{\alpha}_{\alpha UU}\\
	+ s^{\alpha}_{\alpha AU} + s^{\alpha}_{\alpha +-})U^i] + 4\mathbf{e}_{(12)}[\left(s^{i0}_{AA} + s^{i0}_{UU} + s^{i0}_{AU} \right)A_0 + \left(s^{ij}_{AA} + s^{ij}_{UU} + s^{ij}_{AU} \right)U_j + 18(s^{\alpha}_{\alpha AA} + s^{\alpha}_{\alpha UU} + s^{\alpha}_{\alpha AU} +\\
	+ s^{\alpha}_{\alpha +-})A^i] + \sqrt{2}\left(\mathbf{e}_{(23)} + i\mathbf{e}_{(24)}\right)\left(s^{i0}_{-U}V^+_0 + s^{ij}_{-U}V^+_j + 18s^{\alpha}_{\alpha -U}V^{i+}\right) + \sqrt{2}\left(\mathbf{e}_{(23)} - i\mathbf{e}_{(24)}\right)(s^{i0}_{+U}V^-_0 + s^{ij}_{+U}V^-_j\\
	18s^{\alpha}_{\alpha +U}V^{i-}) + \sqrt{2}\left(\mathbf{e}_{(13)} + i\mathbf{e}_{(14)}\right)\left(s^{i0}_{-A}V^+_0 + s^{ij}_{-A}V^+_j + 18s^{\alpha}_{\alpha -A}V^{i+}\right) + \sqrt{2}\left(\mathbf{e}_{(13)} - i\mathbf{e}_{(14)}\right)(s^{i0}_{+A}V^-_0+\\
	+s^{ij}_{+A}V^-_j + 18s^{\alpha}_{\alpha +A}V^{i-}) -\beta_2\left[\mathbf{e}_{(22)}\left(S^{\alpha}_{\alpha U}U^i + 2S^{i0}_U U_0 + S^{ij}_U U_j\right) + \mathbf{e}_{(11)}\left(S^{\alpha}_{\alpha A}A^i + 2S^{i0}_A A_0 +  S^{ij}_A A_j\right)\right]\\
	-\beta_2\mathbf{e}_{(12)}\left[S^{\alpha}_{\alpha U}A^i +S^{\alpha}_{\alpha A}U^i + 2S^{i0}_U A_0 + 2S^{i0}_A U_0 + 2S^{ij}_U A_j + 2 S^{ij}_U A_j \right]\\
	\beta_2\left(\mathbf{e}_{(33)} + \mathbf{e}_{(44)}\right)\left(S^{\alpha}_{\alpha + }V^{i -} + S^{\alpha}_{\alpha -}V^{i+} + 2S^{i0}_+ V^-_0  + 2S^{i0}_- V^+_0 + 2S^{ij}_+ V^-_j + 2S^{ij}_- V^+_j\right)
\end{multline}
	
	For $V_{\mu}^{+}$:
	
	Spin-1:
	\begin{multline}
		\rho^T_+ \equiv + 2i \mathbf{e}_{[34]}\left(\vec{E}_A + \vec{E}_U + \vec{\mathbf{e}}_{+-}\right)\cdot \vec{V}^+ - \sqrt{2}\left(\mathbf{e}_{[13]}- i\mathbf{e}_{[14]}\right)\left(\vec{E}_+ + \vec{\mathbf{e}}_{+A} + \vec{\mathbf{e}}_{+U}\right)\cdot \vec{A}+\\
		-\sqrt{2}\left(\mathbf{e}_{[23]} -i\mathbf{e}_{[24]}\right)\left(\vec{E}_+ + \vec{\mathbf{e}}_{+A} + \vec{\mathbf{e}}_{+U}\right) \cdot \vec{U}+ 
	\end{multline}
and 
	
	\begin{multline}
		\vec{J}^{T}_{+} \equiv 2i\mathbf{e}_{[34]}\left[\left(\vec{E}_{A} + \vec{E}_{B}\right) \cdot \phi^+ + \left(\vec{B}_{A} + \vec{B}_{U}\right) \times \vec{V}^{+} + 2\left( \vec{\mathbf{e}}_{AU} + \vec{\mathbf{e}}_{+-}\right)\cdot \phi^+ + 2 \left(\vec{b}_{AU} + \vec{b}_{+-} \vec{V}^+\right)\right] +\\
		-\sqrt{2}\left( \mathbf{e}_{[13]} -i\mathbf{e}_{[14]} \right)\left[\left(\vec{E}_+ + \vec{\mathbf{e}}_{+A} + \vec{\mathbf{e}}_{+U}\right) \cdot \phi_A + \left(\vec{B}_+ + \vec{b}_{+A} + \vec{b}_{+U} \right) \times \vec{A}\right] \\
		-\sqrt{2}\left( \mathbf{e}_{[23]} -i\mathbf{e}_{[24]} \right)\left[\left(\vec{E}_+ + \vec{\mathbf{e}}_{+A} + \vec{\mathbf{e}}_{+U}\right) \cdot \phi_U + \left(\vec{B}_+ + \vec{b}_{+A} + \vec{b}_{+U} \right) \times \vec{U}\right] 
	\end{multline}

	For $V_{\mu}^{-}$:
	
Spin-1:
\begin{multline}
	\rho^T_- \equiv + 2i \mathbf{e}_{[34]}\left(\vec{E}_A + \vec{E}_U + \vec{\mathbf{e}}_{+-}\right)\cdot \vec{V}^- - \sqrt{2}\left(\mathbf{e}_{[13]}+ i\mathbf{e}_{[14]}\right)\left(\vec{E}_- + \vec{\mathbf{e}}_{-A} + \vec{\mathbf{e}}_{-U}\right)\cdot \vec{A}+\\
	-\sqrt{2}\left(\mathbf{e}_{[23]} -i\mathbf{e}_{[24]}\right)\left(\vec{E}_- + \vec{\mathbf{e}}_{-A} + \vec{\mathbf{e}}_{-U}\right) \cdot \vec{U} 
\end{multline}
and
\begin{multline}
	\vec{J}^{T}_{-} \equiv 2i\mathbf{e}_{[34]}\left[\left(\vec{E}_{A} + \vec{E}_{B}\right) \cdot \phi^- + \left(\vec{B}_{A} + \vec{B}_{U}\right) \times \vec{V}^{-} + 2\left( \vec{\mathbf{e}}_{AU} + \vec{\mathbf{e}}_{+-}\right)\cdot \phi^- + 2 \left(\vec{b}_{AU} + \vec{b}_{+-}\right) \times \vec{V}^-\right] +\\
	-\sqrt{2}\left( \mathbf{e}_{[13]} +i\mathbf{e}_{[14]} \right)\left[\left(\vec{E}_- + \vec{\mathbf{e}}_{-A} + \vec{\mathbf{e}}_{-U}\right) \cdot \phi_A + \left(\vec{B}_- + \vec{b}_{-A} + \vec{b}_{-U} \right) \times \vec{A}\right] \\
	-\sqrt{2}\left( \mathbf{e}_{[23]} +i\mathbf{e}_{[24]} \right)\left[\left(\vec{E}_- + \vec{\mathbf{e}}_{-A} + \vec{\mathbf{e}}_{-U}\right) \cdot \phi_U + \left(\vec{B}_- + \vec{b}_{-A} + \vec{b}_{-U} \right) \times \vec{U}\right] 
\end{multline}

For spin-0:
\begin{multline}
	\rho^{-s}_{V} \equiv \left(\mathbf{e}_{33} + \mathbf{e}_{(44)}\right)\left[\left( \beta_1\vec{S}_{A} + \beta_2 \vec{S}_U\right)\cdot \vec{V}^{-} + \left(\beta_1S^{\alpha}_{\alpha A}  + \beta_2S^{\alpha}_{\alpha U} + 17\rho_1S^{\alpha}_{\alpha A} + \rho_2S^{\alpha}_{\alpha U} \right)\phi^{-}\right]\\
	+\sqrt{2}\left(\mathbf{e}_{(13)} + i \mathbf{e}_{(14)}\right)\left[\beta_- \vec{S}_{-}\cdot \vec{A} +  \left(\beta_- + 17 \rho_- \right)S^{\alpha}_{\alpha -}\phi_A\right] +\sqrt{2}\left(\mathbf{e}_{(23)} + i \mathbf{e}_{(24)}\right)[\beta_- \vec{S}_{-}\cdot \vec{U} +  \\
	+\left(\beta_- + 17 \rho_- \right)S^{\alpha}_{\alpha -}\phi_U] + 2 \left(\mathbf{e}_{(33)} + \mathbf{e}_{(44)}\right)\left[\vec{s}_{+-} \cdot \vec{V}^{-} + \left(s^{\alpha}_{\alpha +-} + 16s^{\alpha}_{\alpha AA} + 16s^{\alpha}_{\alpha UU} \right)\right]\\
	+\sqrt{2}\left(\mathbf{e}_{(13)} + i \mathbf{e}_{(14)}\right)\left[\left(\vec{s}_{-A} + \vec{s}_{-U}\right) \cdot \vec{A} + 17\left(s^{\alpha}_{\alpha -A} + s^{\alpha}_{\alpha -U}\right)\phi_A\right] + 2\left(\mathbf{e}_{(33)} + \mathbf{e}_{(44)}\right)(\vec{s}_{AA} +\\
	+\vec{s}_{UU} + \vec{s}_{AU} + \vec{s}_{+-})\cdot \vec{V}^- + \sqrt{2}\left(\mathbf{e}_{(23)} + i\mathbf{e}_{(24)}\right)\left[\left(\vec{s}_{-A} + \vec{s}_{-U}\right) \cdot \vec{U} + 17\left(s^{\alpha}_{\alpha -A} + s^{\alpha}_{\alpha -U}\right)\phi_U\right] \\
	+ \left(\mathbf{e}_{(33)} - \mathbf{e}_{(44)}\right)\left(\vec{s}_{--} \cdot \vec{V}^{+} + 16s^{\alpha}_{\alpha --}\phi^{+}\right) +2i\mathbf{e}_{(34)}\vec{s}_{--}\cdot \vec{V}^{+} +32\mathbf{e}_{(34)}s^{\alpha}_{\alpha--}\phi^+\\
	-\beta_+\left(\mathbf{e}_{(13)} - i\mathbf{e}_{(14)}\right)\left[\frac{1}{\sqrt{2}}\left(S^{\alpha}_{\alpha-}\phi_A + S^{\alpha}_{\alpha A}\phi^-\right) + \sqrt{2}\left(\vec{S}_{-}\cdot\vec{A} + \vec{S}_{A}\cdot \vec{V}^{-}\right)\right]\\
	-\beta_+\left(\mathbf{e}_{(23)} - i\mathbf{e}_{(24)}\right)\left[\frac{1}{\sqrt{2}}\left(S^{\alpha}_{\alpha-}\phi_U + S^{\alpha}_{\alpha U}\phi^-\right) + \sqrt{2}\left(\vec{S}_{-}\cdot\vec{U} + \vec{S}_{U}\cdot \vec{V}^{-}\right)\right]
\end{multline}
and
\begin{multline}
	j^{i-s}_{V} \equiv \left(\mathbf{e}_{33} + \mathbf{e}_{(44)}\right)\left[\left(\beta_1S_A^{0i} + \beta_2S_U^{0i}\right)V_0{-} + \left(\beta_1S^{\alpha}_{\alpha}\right)V^{i} + \left(\rho_1S^{\alpha}_{\alpha A} + \rho_2S^{\alpha}_{\alpha U}\right)V^{i-}\right]+\\
	+\sqrt{2}\left(\mathbf{e}_{(13)} + i \mathbf{e}_{(14)}\right)\left[\beta_-S^{0i-}A_0 + \beta_-S^{ji}A_j + \left(\beta_- + 17 \rho_-\right)S^{\alpha -}_{\alpha}A^{i}\right] +\\
	+\sqrt{2}\left(\mathbf{e}_{(23)} + i \mathbf{e}_{(24)}\right)\left[\beta_-S^{0i-}U_0 + \beta_-S^{ji}U_j + \left(\beta_- + 17 \rho_-\right)S^{\alpha -}_{\alpha}U^{i}\right] +\\
	+\left(\mathbf{e}_{33} + \mathbf{e}_{(44)}\right)\left[\left(s^{0i}_{AA} + s^{0i}_{UU} + s^{0i}_{AU}\right)V_0^{-} + \left(s^{ji}_{AA} + s^{ji}_{UU} + s^{ji}_{AU}\right)V_j^{-}+8\sqrt{2}\left(s^{\alpha}_{\alpha AA} + s^{\alpha}_{\alpha UU}\right)V^{i-} \right]+\\
	+\sqrt{2}\left(\mathbf{e}_{(13)} + i\mathbf{e}_{(14)}\right)\left[\left(s^{0i}_{-A} + s^{0i}_{-U}\right)A_0 + \left(s^{ji}_{-A} +s^{ji}_{-U}\right)A_{j} + \left(s^{\alpha}_{\alpha -A} + s^{\alpha}_{\alpha -U}\right)A^i\right]+\\
	+\sqrt{2}\left(\mathbf{e}_{(23)} + i\mathbf{e}_{(24)}\right)\left[\left(s^{0i}_{-A} + s^{0i}_{-U}\right)U_0 + \left(s^{ji}_{-A} +s^{ji}_{-U}\right)U_{j} + \left(s^{\alpha}_{\alpha -A} + s^{\alpha}_{\alpha -U}\right)U^i\right]+\\
	\left(\mathbf{e}_{33} - \mathbf{e}_{(44)}\right)\left(s^{0i}_{--}V^{0+} + s^{ji}_{--}V_i^+ + 16s^{\alpha}_{\alpha --}V^{i}\right) 2i\mathbf{e}_{(34)}\left(s^0i_{--}V_0^+ + s^{ji}_{--}V^{ji}V_j^+ + 16s^{\alpha}_{\alpha --}V^{i+}\right)\\
	-\beta_+\left(\mathbf{e}_{(13)} - i\mathbf{e}_{(14)}\right)\left\{\frac{1}{\sqrt{2}}\left[S^{\alpha-}_{\alpha}A^{i} + S^{\alpha}_{\alpha A}V^{i-}\right]+\sqrt{2}\left[S_-^{0i}A_0 + S_-^{ji}A_j + S^{0i}_AV_0^{-} + S^{ji}_A V^{i-}\right]\right\}\\
	-\beta_+\left(\mathbf{e}_{(23)} - i\mathbf{e}_{(24)}\right)\left\{\frac{1}{\sqrt{2}}\left[S^{\alpha-}_{\alpha}U^{i} + S^{\alpha}_{\alpha U}V^{i-}\right]+\sqrt{2}\left[S_-^{0i}U_0 + S_-^{ji}U_j + S^{0i}_UV_0^{-} + S^{ji}_U V^{i-}\right]\right\}
\end{multline}

\end{appendices}

\end{document}